\documentclass[10pt]{article}
\scrollmode
\usepackage{amsmath,amssymb}
\usepackage{eucal}
\usepackage[dviwindo]{graphicx}
\usepackage[dviwindo]{graphics}

\numberwithin{equation}{section}
\newcommand{\ben}{\begin{eqnarray}}
\newcommand{\een}{\end{eqnarray}}
\newcommand{\la}{\label}

\begin{document}

\title{Classes of Exact Solutions to the Teukolsky  Master Equation}

\vskip 1.5truecm

\author{P.~P.~Fiziev\thanks{Department of Theoretical Physics, University of Sofia,
Boulevard 5 James Bourchier, Sofia 1164, Bulgaria,
E-mail:\,\,\,fiziev@phys.uni-sofia.bg,
and BLTF, JINR, Dubna, 141980 Moscow Region, Rusia,
E-mail:\,\,\,fiziev@theor.jinr.ru}
}

\date{}
\maketitle

\begin{abstract}

The Teukolsky Master Equation is the basic tool for study of perturbations of the Kerr metric in linear approximation.
It admits separation of variables, thus yielding the Teukolsky Radial Equation and the Teukolsky Angular Equation.
We present here a unified description of all classes of exact solutions to these equations
in terms of the confluent Heun functions.
Large classes of new exact solutions are found and classified with respect to their characteristic properties.
Special attention is paid to the polynomial solutions which are singular ones and
introduce collimated one-way-running waves. It is shown that a proper linear combination of such
solutions can present bounded one-way-running waves. This type of waves
may be suitable as models of the observed astrophysical jets.
\end{abstract}

%
\sloppy
\scrollmode

\section{Introduction}
At present the study of different types of perturbations of the gravitational field of
black holes, neutron stars and other compact astrophysical objects
is a very active field for analytical, numerical, experimental and
astrophysical research. Ongoing and nearest future experiments
based on perturbative and/or numerical analysis of relativistic
gravitational dynamics are expected to provide critical tests of the
existing theories of gravity \cite{KipTorn}.

The study of perturbations of rotating relativistic objects in Einstein GR was
pioneered by Teukolsky \cite{Teukolsky} by making use of the famous
Teukolsky Master Equation (TME). It describes the perturbations
${}_s\Psi(t,r,\theta,\varphi)$ of all physically interesting spin-weights $s=0,
\pm 1/2, \pm 1, \pm 3/2, \pm 2$
to the Kerr background metric in terms of the corresponding Newman-Penrose scalars.
The pairs of spin-weights $s$ with opposite signs $\sigma=\text{sign}(s)=\pm 1$
correspond to two different perturbations with opposite helicity and spin  $|s|=0, 1/2, 1, 3/2$,
or $2$.
Under proper boundary conditions
for the TME one obtains quasi-normal modes (QNM) of the Kerr black holes. Various significant results and
additional references can be found in \cite{Chandra}-\cite{QNM}.

The key feature of the TME is that in the Boyer-Lindquist coordinates
one can separate the variables using the ansatz
$\Psi(t,r,\theta,\varphi)= e^{-i\omega t} e^{im \varphi}S(\theta)R(r)$,
i.e. looking for solutions in a specific factorized form. Thus, one obtains a pair of two connected
ordinary differential equations for the nontrivial factors ${}_sS_{\omega,E,m}(\theta)$
and ${}_sR_{\omega,E,m}(r)$ -- the Teukolsky
angular equation (TAE) \cite{Teukolsky,Chandra,angular}
\begin{subequations}\la{angE:ab}
\ben
{\frac {1}{\sin \theta}} {\frac {d}{d\theta}} \left( \sin
\theta {\frac {d} {d\theta}}\,{}_sS_{\omega,E,m}( \theta ) \right)
+{}_sW_{\omega,E,m}(\theta) {}_sS_{\omega,E,m}(\theta)\!=0, \hskip 2.2truecm \la{angE:a}\\
{}_sW_{\omega,E,m}(\theta)=E+a^2\omega^2\cos^{2}\theta-2sa\omega \cos
\theta -(m^2+s^2+2ms\cos\theta)/\sin^2 \theta; \la{angE:b}
\een
\end{subequations}
and the Teukolsky radial equation (TRE) \cite{Teukolsky,Chandra}

\begin{subequations}\la{radE:ab}
\ben
{\Delta}^{-s}{\frac {d }{dr}} \left( {\Delta}^{s+1}
{\frac{d}{dr}}\,{}_sR_{\omega,E,m}(r)  \right) + {}_sV_{\omega,E,m}(r)\, {}_sR_{\omega,E,m}(r) =0, \hskip 0truecm
\la{radE:a}\\
{}_sV_{\omega,E,m}(r)={\frac 1 \Delta}{{K}^{2}} -is {\frac 1 \Delta} {\frac {d\Delta} {dr}}K-L.\hskip 2truecm
\la{radE:b}
\een
\end{subequations}
The azimuthal number may have values $m=0, \pm 1, \pm 2,\dots $
for integer spin, or $m= \pm 1/2, \pm 3/2,\dots $ for half-integer spin \cite{HalfInteger}.
In Eq. \eqref{radE:ab} we use the expressions
$\Delta\!=\!r^2\!-\!2Mr\!+\!a^2$, $K\!=\!\omega(r^2\!+\!a^2)\!-\!ma$,
$L\!=\!E\!-\!s(s\!+\!1)\!+\!a^2\omega^2\!-\!2ma\omega\!-\!4\,is\omega\,r$.
The real parameter $a\!=\!J/M$ is related with the angular momentum $J$ of the Kerr metric,
$M$ being its Keplerian mass.
The two {\em complex} parameters $\omega$ and $E$ -- the constants of separation,
are to be determined using the boundary conditions of the problem.

The negativity of the imaginary part
$\omega_I=\Im(\omega)<0$ of the complex frequency $\omega=\omega_R+i\omega_I$
ensures linear stability of the solutions in the exterior domain of the Kerr metric
with respect to the future time direction $t\to +\infty$ \cite{Teukolsky, Kerr_Stability}.
In the interior domain the solutions to the TME are not stable \cite{Kerr_Instability}.

From a mathematical point of view the function
${}_s{\mathcal K}_{\omega,E,m}(t,r,\theta,\varphi)\sim e^{-i\omega t} e^{im \varphi}\,{}_sS_{\omega,E,m}(\theta)\,{}_sR_{\omega,E,m}(r)$
actually defines a factorized {\em kernel} of the general integral representation  for the solutions to the TME:
\begin{equation}\la{IntRepresentation_E_omega}
{}_s\Psi(t,r,\theta,\varphi)= {1\over 2\pi} \int\!\!d\omega\int\!\!dE\,
\sum_m
{}_sA_{\omega,E,m}\,e^{-i\omega t}\, e^{im\varphi}\,{}_sS_{\omega,E,m}(\theta)\,{}_sR_{\omega,E,m}(r).
\end{equation}

The formal mathematical representation \eqref{IntRepresentation_E_omega} is written ad hoc
as the most general superposition of all particular solutions.
In it a summation on all admissible values of the two separation constants $\omega$ and $E$ is assumed.
Its usefulness will be illustrated by different examples in what follows.

It is well known \cite{Carter} that the Carter separation constant (which is equivalent to the constant $E$, used here)
may be related with the total angular momentum of the solution ${}_s\Psi(t,r,\theta,\varphi)$.
Under proper boundary conditions for the TAE this momentum has discrete values defined
by an (half)integer $l$ \cite{Teukolsky}. If we are interested in superpositions
${}_s\Psi(t,r,\theta,\varphi)$ of solutions with a definite total angular momentum,
the integration with respect to the constant $E$ must be replaced with summation over the (half)integer $l$.
Thus, instead of the most general linear mixture \eqref{IntRepresentation_E_omega}
we have to use the representation of the solutions
\begin{equation}\la{IntRepresentation_l}
{}_s\Psi(t,r,\theta,\varphi)= {1\over 2\pi} \int\!\!d\omega\sum_{l}\,\sum_m
{}_sA_{\omega,l,m}\,e^{-i\omega t}\, e^{im\varphi}\,{}_sS_{\omega,l,m}(\theta)\,{}_sR_{\omega,l,m}(r),
\end{equation}
introduced in the problem at hand for the first time in \cite{Teukolsky}.
The transition from the representation \eqref{IntRepresentation_E_omega} to
representation \eqref{IntRepresentation} is formally equivalent to the use
of a {\em  singular} kernel proportional to the sum of Dirac $\delta$-functions:
$\sum_{l}\delta(E-{}_sE(\omega,l,m))$
in \eqref{IntRepresentation_E_omega}.
Here ${}_sE(\omega,l,m)$ belongs to some spectrum which is specific for the given problem and
is completely defined by the corresponding boundary conditions, see section 7.

Further on, the boundary conditions may fix some discrete spectrum
for the frequencies $\omega$ in \eqref{IntRepresentation}. Then the integral on $\omega$ will be replaced
by discrete summation over some $\omega_n$.
This is equivalent to the use once more of a singular kernel, now proportional to
$\sum_n \delta (\omega-\omega_n)$. As a result one obtains
\begin{equation}\la{IntRepresentation}
{}_s\Psi(t,r,\theta,\varphi)= \sum_{n}\,\sum_{l}\,\sum_m
{}_sA_{n,l,m}\,e^{-i\omega_n t}\, e^{im\varphi}\,{}_sS_{n,l,m}(\theta)\,{}_sR_{n,l,m}(r).
\end{equation}

The lack of a rigorous mathematical theory explaining the Teukolsky separation is an old and well-known
problem\footnote{The author is grateful to unknown referee for this important remark.}.
The study of the QNM \cite{QNM} not only illustrates the above situation but also shows that the kernel
${}_s{\mathcal K}_{\omega,E,m}(t,r,\theta,\varphi)$ can be singular with respect
to the variable  $r$ at infinity and at the horizons.
In the existing literature only regular with respect to the variable $\theta$
kernels ${}_s{\mathcal K}_{\omega,E,m}(t,r,\theta,\varphi)$ are in use.
In the present paper, we start the consideration of both regular and singular with respect
to the angle $\theta$ kernels in the integral representation \eqref{IntRepresentation_E_omega}
of the solutions to the TME.
Different types of kernels are to be used for solution of different boundary problems.
Note that from a physical point of view the regularity of the very solution
${}_s\Psi(t,r,\theta,\varphi)$ in equations \eqref{IntRepresentation_E_omega}, \eqref{IntRepresentation_l}
and \eqref{IntRepresentation} is important.
The kernels like ${}_s{\mathcal K}_{\omega,E,m}(t,r,\theta,\varphi)$
are auxiliary mathematical objects. One is often forced to use singular kernels
in the natural integral representations of the solution to physical problems.
The regularity of the very physical solution ${}_s\Psi(t,r,\theta,\varphi)$
with respect to the variable $\theta$ depends on the choice of the amplitudes ${}_sA_{\omega,E,m}$.
It can be guaranteed by a suitable choice of these amplitudes, as shown in section 10.

Despite the essential progress both in the numerical study \cite{KerrQNM_numeric} of the
solutions to   equations (\ref{angE:a}) and (\ref{radE:a})
and in the investigation of their analytical properties \cite{KerrQNM_analytic},
at present there exists a number of basic questions remaining unanswered.
For example, it has been well known for a long time \cite{TME_Heun} that
the TAE (\ref{angE:a}) and TRE (\ref{radE:a}) can be reduced to the confluent Heun ordinary
differential equation \cite{Heun} written here in the following simplest uniform shape
\cite{Fiziev2009a, Fiziev2009b}:
\begin{align}
H''+\left(\alpha+{{\beta+1}\over{z}}+{{\gamma+1}\over{z-1}}\right)H'+
\left( {\mu\over z}+{\nu\over{z-1}} \right)H = 0.\la{DHeunC}
\end{align}

The constants $\mu$ and $\nu$ in Eq. \eqref{DHeunC} are related with the constants $\alpha, \beta, \gamma, \delta, \eta $,
accepted in the notation $\text{HeunC}(\alpha,\beta,\gamma,\delta,\eta,z)$ as follows:
\begin{subequations}\la{munu:ab}
\ben
\delta =\mu+\nu - \alpha\,{\frac{\beta+\gamma+2} 2},\la{munu:a}\\
{\eta={\frac{\alpha(\beta+1)} 2}-\mu-\frac{\beta +\gamma+\beta\gamma} 2}.\la{munu:b}
\een
\end{subequations}

To the best of our knowledge we still do not have a detailed description of the exact analytical solutions to
the TAE \eqref{angE:a} and  TRE \eqref{radE:a} in terms of the {\em confluent Heun function}
$\text{HeunC}(\alpha,\beta,\gamma,\delta,\eta,z)$ -- the unique particular local
solution of Eq. \eqref{DHeunC} which is regular in the vicinity of
the regular singular point $z\!=\!0$ and obeys the normalization condition
$\text{HeunC}(\alpha,\beta,\gamma,\delta,\eta,0)\!=\!1$ \cite{Heun}.\footnote{
In the present paper, we use the Maple-computer-package notation for the Heun functions.
Basically, this notation is borrowed from the two mile-stone papers
on modern theory of Heun's functions by Decarreau et al. in \cite{Heun}; at present it seems to be
most popular, since the Maple package is the only one for analytical and numerical
work with Heun's functions.}.
Note that other particular solutions to equation (\ref{DHeunC}), as well as its
general solution, are {\em not} termed "confluent Heun's functions", according to the accepted modern terminology \cite{Heun}.
The reason is that, in general, other solutions can be represented in a nontrivial way in terms of solutions
$\text{HeunC}(\alpha,\beta,\gamma,\delta,\eta,z)$ of the corresponding arguments.
Hence, from a computational point of view it is sufficient to study
{\em only} the Taylor series  of this standard local solution
and its analytical continuation in the complex plane $\mathbb{C}_z$.
Thus, the instrumental use of the confluent Heun function $\text{HeunC}(\alpha,\beta,\gamma,\delta,\eta,z)$ -- the basic purpose
of the present paper,
is much more advantageous than the simple fact, recognized already in \cite{TME_Heun},
that the TRE and  TAE can be reduced to the confluent Heun equation \eqref{DHeunC}.

In the late 2006 a program for filling the above gaps in the study of the TME was started
as a natural extension of the papers \cite{F},
where a similar approach was developed for the Regge-Wheeler equation (RWE).
The first results were quite stimulating \cite{PFDS},
but serious difficulties came across in both analytical and numerical studies.
This is because the theory of Heun's functions,
as well as numerical tools for calculations with them still are not developed enough.

Here we pay special attention to the polynomial solutions of Eq.  (\ref{DHeunC}).
According to \cite{Heun}, the confluent Heun function $\text{HeunC}(\alpha,\beta,\gamma,\delta,\eta,z)$
reduces to a polynomial of degree $N\geq 0$ of the variable $z$,
if and only if the following two conditions are satisfied:
\begin{subequations}
\la{PolynomCond:ab}
\ben
{\frac{\delta}{\alpha}}+{\frac{\beta+\gamma}{2}}+N+1=0,\la{PolynomCond:a}\\
\Delta_{N+1}(\mu)=0.\la{PolynomCond:b}
\een
\end{subequations}
We call the first condition \eqref{PolynomCond:a}
a "$\delta_N$-condition", and the second one \eqref{PolynomCond:b} -- a "$\Delta_{N+1}$-condition".
An explicit form of the "$\Delta_{N+1}$-condition" in form of a determinant useful for practical calculations,
as well as a novel derivation of confluent Heun's polynomials can be found in \cite{Fiziev2009a}.
A recurrent procedure for calculation of $\Delta_{N+1}$  (\ref{PolynomCond:b}) and its relation with
Starobinsky's constants are presented in \cite{Fiziev2009b}.

On the other hand, the so-called algebraically special solutions to the RWE and the TRE
were discovered long time ago \cite{AlgSpetial}.
These are of a generalized polynomial type,
i.e. products of polynomials and simple non-polynomial factors
which are elementary functions.
According to the existing literature,
these solutions describe pure incoming or pure outgoing waves.
The algebraically special solutions still are not discussed in terms of Heun's polynomials.
To our knowledge, attempts for application of this class of solutions to real physical problems
cannot be found in the existing literature on gravitational physics.

The only exception are the recent papers \cite{PFDS, PFDS_astroph:HE},
where one can find some preliminary results.
There, special polynomial solutions of the TAE  were considered in more detail.
In particular, in the first two of the articles \cite{PFDS_astroph:HE}
it was  demonstrated that these singular with respect to the angular variable $\theta$ solutions can describe
collimated waves which resemble the observed astrophysical jets.
In the third of these papers the spectrum of
electromagnetic jets from Kerr black holes and naked singularities in the Teukolsky
perturbation theory was calculated for the first time,
using some of the basic results of the present paper.

Very recently the algebraically special solutions of the RWE and the TME were proved to be
relevant for the study of instabilities of different kind of some more or less "exotic"
solutions to the Einstein equations \cite{ASM_non_stabilities}.
Physical manifestation of the instabilities of the mathematical solutions
are the explosions of the corresponding objects.
Therefore, it seems natural to look for a perturbative description
of explosions in terms of solutions of the TME, which are stable in the future and
instable in the past.
The confluent Heun functions give a rigorous mathematical basis for analysis of these problems.

On the other hand, the recently found properties of confluent Heun's function \cite{Fiziev2009a}
show that one can introduce a new subclass of "$\delta_N$-confluent-Heun's-functions",
which obey {\em only} the $\delta_N$-condition -- Eq. \eqref{PolynomCond:a}.
In \cite{Fiziev2009b} is shown that such "$\delta_N$-solutions" of the TRE and the TAE define the most general class
of solutions, for which properly generalized Teukolsky-Starobinsky's identities exist. Moreover, this approach
reveals the existence of Teukolsky-Starobinsky's type of identities for Regge-Wheeler and Zirilli equations, as well.
Here we study in more detail the $\delta_N$-solutions to the TRE and  TAE.
In particular, we show that the regular solutions to the TAE,
which are the only class of solutions to the TAE, used up to now \cite{Teukolsky, Chandra,angular},
are precisely nonpolynomial $\delta_N$-solutions.
In contrast, the polynomial solutions to the TAE of all spins are shown to be singular
around one of the poles ($\theta=0$, or $\theta=\pi$) of the unit sphere $\mathbb{S}^{\,(2)}_{\theta,\varphi}$
and regular around the other one.
This new situation reflects the specific properties of the confluent Heun function.
It is not consistent with our experience, based on the work with hypergeometric functions, solving
the angular part of the Laplace equation in celestial and quantum mechanics, or in electrostatics.
It is well known that in the last case solutions regular in the interval $\theta \in [0,\pi]$
(including both the poles) are polynomial.

In the limit $a \to 0$, when the Kerr metric approaches the non-rotating Schwarzschild one,
there exist a smooth transition  from perturbations
of the Kerr metric to perturbations of the Schwarzschild metric in terms of the Weyl scalars,
but a simple transition from the solutions of the TME to the solutions of the RWE
(see \cite{Fiziev2009b,F}) is not possible \cite{Chandra}.
Nevertheless, the mathematical analogy between the corresponding solutions becomes quite transparent
when the solutions are represented in terms of the confluent Heun functions \cite{Fiziev2009b,F}.
The limit $a \to 0$  is traced in more detail in section 4.1.2 -- for the TRE and in section 7 -- for the TAE.

This way, using confluent Heun's function, we hope to obtain a more clear picture
of the quite complicated present-day state of the arts in the perturbation
theory under consideration and its possible further developments and new physical applications.

The main purposes of the present paper is to report some of the basic results,
obtained for a detailed description of the exact solutions of the TME
in terms of the confluent Heun function $\text{HeunC}(\alpha,\beta,\gamma,\delta,\eta,z)$,
to introduce a large number of new classes of such solutions,
and to formulate some interesting boundary problems for the TRE, TAE,
and TME in terms of confluent Heun's functions.

Besides the already stressed new developments, in the present paper for the first time we
introduce and study the differential invariants of the Weyl tensor,
which indicate in an invariant way both the event and Cauchy horizons of the Kerr metric as singular points of the TRE (section 2.1),
the explicit form of 16 classes of exact solutions to the TRE in terms of confluent Heun's functions (section 2.2),
a new classification of the solution to the TRE, based on specific properties of confluent Heun's functions (section 3),
especially, the class of $\delta_N$-radial solutions and, in particular,
two unknown infinite classes of exact solutions with equidistant complex spectra of frequencies,
two novel classes of polynomial solutions to the TRE (section 4),
the explicit form of 16 classes of exact solutions to the TAE in terms of confluent Heun's functions (section 5),
a new concomitant confluent Heun's function and its application to the TAE (section 5),
a new classification of the solution to the TAE, based on specific properties of confluent Heun's functions (section 6),
especially, the class of $\delta_N$-angular solutions,
a novel description of the regular solutions to the TAE in terms of confluent Heun's functions (section 7),
two classes of singular polynomial solutions to the TAE (section 8),
256 classes of exact factorized solutions to the TME (section 9),
an explicit construction of exact bounded solutions to the TME with spin $1/2$,
using the singular kernel, built from the polynomial solutions to the TAE (section 10),
and novel general exact solutions of the TME in the form of one-way running waves (section 10).
It seems natural from physical point of view to use these solutions for study of the relativistic jets.
Some other general conclusions and perspectives for further developments are outlined in the
concluding section 11.

\section{Exact Solutions to the Teukolsky Radial Equation in Terms of the Confluent Heun functions}

\subsection{Explicit form of the TRE and Geometrical Character of its Singularities}

Much like in the case of the Schwarzschild solution, for the Kerr one we have a complicated space-time structure
and a different physical meaning of the space-time coordinates in the different domains.
For example, consider, as usual, only the real values of $r$.
In the interior of the Kerr metric: $0\leq r_{-}<r< r_{+}$ -- between the zeros $r_{\pm}=M\pm\sqrt{M^2-a^2}, a\leq M$
of the function $\Delta$ (i.e., between the Cauchy horizon $r_{-}$
and the event horizon $r_{+}$),
two of the eigenvalues: $\lambda_{t}$ and $\lambda_{r}$ of the metric in the Boyer-Lindquist
coordinates simultaneously change their signs. Indeed, one pair of eigenvalues is
$\lambda_{\theta}=g_{\theta\theta}=r^2+a^2\cos^2\theta$ and $\lambda_{r}=g_{rr}=(r^2+a^2\cos^2\theta)/\Delta$.
The second pair of eigenvalues is the roots $\lambda_{t}, \lambda_{\phi}$ of the equation
$\lambda^2-(g_{tt}+g_{\phi\phi})\lambda+g_{tt}g_{\phi\phi}-g_{t\phi}^2$.
Their product equals $\lambda_{t}\lambda_{\phi}=-\Delta\sin^2\theta$.
The last expression, together with the form of  $g_{rr}$ proves the simultaneous change of the signs of the
two eigenvalues $\lambda_{t}, \lambda_{r}$, when the variable $r$ crosses the horizons $r_{\pm}$,
since the determinant  $g=-(r^2+a^2\cos^2\theta)^2\sin^2\theta$ of the metric does not vanish there.
As a result, between the two horizons $r_{\pm}$ the variable $t_{in}=x\in (-\infty,\infty)$
plays the role of the interior time and the variable $r_{in}=t$ is the interior radial variable.
We use the following  Kerr-metric-tortoise-coordinate:
$r_{\!*}=r+a_{+}\ln|(r-r_{+})/(r_{+}-r_{-})|-a_{-}\ln|(r-r_{-})/(r_{+}-r_{-})|\in (-\infty,\infty)$,
where $a_\pm={\frac{r_{+}+r_{-}}{r_{+}-r_{-}}}r_{\pm}$.
It is a straightforward generalization of the
tortoise variable for the exterior domain $r\in (r_{+},\infty)$ proposed in \cite{Teukolsky}.
Since our expression is valid in the interior domains, too,
the inverse function defines $r=r(t_{in})$ when $r\in (r_{-}, r_{+})$.
In the second interior domain $r<r_{-}$ the variables $r$ and $t$ restore their original meaning.
For a detailed analysis of the light cones in the Kerr geometry see \cite{LC}.
This consideration is necessary for understanding of the physical meaning
of the solutions to the TME in the different Kerr-space-time domains.

The explicit form of the TRE
\ben
\la{TRE}
{\frac {d^{2}R_{\omega,E,m}}{d{r}^{2}}} + (1+s)  \left( {\frac{1}{r-{\it r_{+}}}}+
{\frac{1}{r-{\it r_{-}}}} \right){\frac {dR_{\omega,E,m}}{dr}} + \nonumber\\                                                                                                 +\left( {\frac { \Big( \omega\, \left( {a}^{2}+{r}^{2} \right) -am \Big) ^{2}}{ \left( r-r_{+} \right)  \left( r-r_{-} \right) }}-                             is \left( {\frac{1}{r-{\it r_{+}}}}+ {\frac{1}{r-{\it r_{-}}}} \right)  \Big( \omega\, \left( {a}^{2}+{r}^{2} \right) -am \Big)- \right.\nonumber\\
\left.-E+s(s+1)-{a}^{2}{\omega}^{2}+2m\,a\omega + 4\,i s \omega r  \vphantom{\frac{\Big( a^2\Big)}{\big( a^2 \big)}}\!\right)
{\frac {R_{\omega,E,m}} {( r-r_{+})( r-r_{-})}}=0
\een
shows that it has three singular points: $r=r_{\pm}$ and $r=\infty$.
In the present paper, we consider only the {\em non-extremal} Kerr metric with {\em real} $r_{+}>r_{-}\geq 0$.
Then the first two are regular singular points,
and the third one (the physical infinity $r=\infty$) is an irregular singular point.
The symmetry of Eq. \eqref{TRE} under the interchange $r_{+}\!\leftrightarrows r_{-}$ is obvious.
Thus, we see that the two horizons of the Kerr metric are singularities for the TRE which are to be
treated on equal footing. Do these singularities have an invariant meaning independent of the coordinate choice?

It is well known that the algebraic invariants of the Riemann curvature tensor ${\cal R}_{ijkl}$
are not able to indicate the horizons of the Kerr black hole
and one usually considers them as pure coordinate singularities of the metric in the Boyer-Lindquist coordinates.
In contrast, the circle $r=0, \theta = \pi/2$ is a singularity of the algebraic invariants of the Riemann tensor \cite{Chandra}.
Since the pure {\em algebraic} invariants of the tensor ${\cal R}_{ijkl}$ do not fix completely the geometry,
their consideration is not sufficient to recover all gauge-invariant space-time properties.
For this purpose one must consider a large enough number of high-order-differential-invariants
of the Riemann tensor \cite{Cartan}.

It is not difficult to find differential invariants of the Riemann tensor of the Kerr metric
which are able to distinguish both the horizons $r_{\pm}$ and the ergo-surface $g_{tt}=0$.
Indeed, let us consider the following algebraic invariants of the Weyl tensor ${\cal W}_{ijkl}$:
$I_{1}={\frac 1 {48}}{\cal W}_{ijkl}{\cal W}^{ijkl}$ -- the density of the Euler characteristic class,
and
$I_{2}={\frac 1 {48}}{\cal W}_{ijkl}\,{}^*{\cal W}^{ijkl}$ --
the density of the Chern-Pontryagin characteristic class \cite{char_classes}.
Let us put $(I_{1}-iI_{2})^{1/2}=\lambda=|\lambda|\exp(i\psi)$.
Then $r=\left({\frac M{|\lambda|}}\right)^{1/6}\cos(\psi/6)$ and
$\rho=\left({\frac M{|\lambda|}}\right)^{1/6}\cos(\psi/6)^{-1}$ are obviously invariants of the Weyl tensor
-- nonalgebraic and nondifferential ones.
In the Boyer-Lindquist coordinates one obtains $\rho=r+{\frac{a^2}r}\cos\theta$ and $g_{tt}=1-2M/\rho$.
The differential invariants of first order
\begin{subequations}
\label{KBH_D_inv:ab}
\ben
DI_{1}=-\left(\nabla \ln r\right)^2={\frac{1}{r\rho}}\left(1-{\frac{2M}{r}}+{\frac{a^2}{r^2}}\right),\label{KBH_D_inv:a}\\
DI_{2}=\left(\nabla \ln \rho\right)^2-\left(\nabla \ln r\right)^2=
{\frac 4 {\rho^2}}\left({\frac \rho r}-1\right)\left(1-{\frac{2M}\rho}\right)\label{KBH_D_inv:b}
\een
\end{subequations}
indicate the two Kerr black hole horizons, the ergo-surface and some other geometrical objects in the Kerr space-time.
Thus, the two horizons $r_{\pm}$ of the Kerr metric are shown to be invariant objects, being singularities
of the same kind in equation \eqref{TRE}.

In the limit $a\to 0$ we have $\rho\to r$ and the differential invariant in equation \eqref{KBH_D_inv:b} becomes trivial: $DI_{2}\to 0$.
In the same limit,  the differential invariant \eqref{KBH_D_inv:a} produces a nontrivial result
$\left(1-2M/r\right)/r^2$ for the Schwarzschild metric
which is similar to the one derived already in the third of the papers \cite{Cartan}, as well as in the very recent seventh one.
Differential invariants similar to (2.2) are considered in the sixth of the references [23] without any application.

\subsection{Explicit Form of the Local Solutions to the TRE}

The analytical study of the solutions to the TRE and  TAE was started in \cite{Leaver}
and continued by different approximate methods \cite{QNM, KerrQNM_analytic} without utilizing of Heun's functions.
Using the confluent Heun function one can write down 16 exact local Frobenius type solutions to the TRE (\ref{TRE}) in the form:
\ben
{}_sR^\pm_{\omega,E,m,\sigma_\alpha,\sigma_\beta,\sigma_\gamma}(r;r_{+},r_{-})\Delta^{s/2}\!=\!e^{\sigma_\alpha{\frac{\alpha_{{}_\pm} z_{{}_\pm}} 2}} z_{{}_\pm}^{\sigma_\beta{\frac{\beta_{{}_\pm}} 2}} z_{{}_\mp}^{\sigma_\gamma{\frac{\gamma_{{}_\pm}} 2}}
\text{HeunC}({\sigma_\alpha\alpha_{{}_\pm},\sigma_\beta\beta_{{}_\pm},\sigma_\gamma\gamma_{{}_\pm},\delta_{{}_\pm},\eta_{{}_\pm},z_{{}_\pm}}),
\la{TRE16local}\een
which is very similar to the form of the solutions to the RWE \cite{Fiziev2009b,F}.
Here\footnote{Note that the notation $z_{\pm}$
in Eq. \eqref{TREparameters:f} is consistent with
the limits $z_{\pm}\to \pm \infty$ for $r\to\infty$.
Their relation with the notation of the parameters of the Kerr metric $r_{\pm}$ is illustrated by the equations $z_{\pm}(r_{\mp};r_{+},r_{-})=0$.
The labels $\pm$ in the notation $R^{\pm}$ in Eq. \eqref{TRE16local} are related with the labels of their arguments  $z_{\pm}$,
not with the labels of the parameters $r_{\pm}$.}
\begin{subequations}\label{TREparameters:abcdef}
\begin{align}
\alpha_{{}_{+}}\!&=\!{}_{s}\alpha_{\omega,E,m}(r_{+},r_{-})\!=\!2i\omega(r_{+}-r_{-})\!=i 2 p\, {{\omega}/{\Omega}_a}, \label{TREparameters:a}\\
\beta_{{}_{+}}&=\!{}_{s}\beta_{\omega,E,m}(r_{+},r_{-})\!=s +  i \left(m-\omega/\Omega_{-}\right)/p,\label{TREparameters:b}\\
\gamma_{{}_{+}}&=\!{}_{s}\gamma_{\omega,E,m}(r_{+},r_{-})\!=s -  i \left(m-\omega/\Omega_{+}\right)/p, \label{TREparameters:c}\\
\delta_{{}_{+}}&=\!{}_{s}\delta_{\omega,E,m}(r_{+},r_{-})\!=\alpha_{+}\left(s-i\omega(r_{+}+r_{-})\right)\!=
\alpha_{+}\left(s-i\omega/\Omega_{g}\right),\label{TREparameters:d}\\
\eta_{{}_{+}}\!&=\!{}_{s}\eta_{\omega,E,m}(r_{+},r_{-})\!=\!-E+s^2\!+m^2\!+
{\frac{2m^2\Omega_a^2-\omega^2}{4 p^2\Omega_a^2}}-{\frac{(2m\Omega_a-\omega)^2} {4 p^2\Omega_g^2}}
-{\frac 1 2}\left({s\!-\!i\,\frac{\omega\,\Omega_{+}}{\Omega_a\Omega_g}}\right)^2;\label{TREparameters:e}\\
z_{+}\!&=\!z_{+}(r;r_{+},r_{-})\!=\!{\frac{r\!-\!r_{-}}{r_{+}\!-\!r_{-}}},\,z_{-}\!=\!z_{-}(r;r_{+},r_{-})\!=\!{\frac{r_{+}\!-\!r}{r_{+}\!-\!r_{-}}};\,
z_{+}\!+\!z_{-}\!=\!1,\,z_{+}z_{-}\!=\!{\frac{-\Delta}{(r_{+}\!-\!r_{-})^2}}.\label{TREparameters:f}
\end{align}
\end{subequations}

The discrete parameters $\sigma_\alpha,\sigma_\beta,\sigma_\gamma$
have values $\pm1$\footnote{Further on, $\sigma_x\!=\!\text{sign}(x)$
denotes the sign of the {\em real} quantity $x$. The only exception is $\sigma\!\equiv\!\sigma_s$ where we skip the index $s$.}.
In equations \eqref{TREparameters:a}-\eqref{TREparameters:e} we use the following quantities:
the angular velocity of the event horizon $\Omega_{+}=a/2Mr_{+}={\sqrt{r_{-}/r_{+}}}\big/\left(r_{+}+r_{-}\right)$,
the angular velocity of the Cauchy horizon $\Omega_{-}=a/2Mr_{-}={\sqrt{r_{+}/r_{-}}}\big/\left(r_{+}+r_{-}\right)$,
the arithmetically-averaged angular velocity $\Omega_a=\left(\Omega_{+}+\Omega_{-}\right)/2=1/(2a)$,
the geometrically-averaged angular velocity $\Omega_g=\sqrt{\Omega_{+}\Omega_{-}}=1/(2M)$, and the new dimensionless parameter
\ben\la{p}
p\!=\!{\frac 1 2}\left(\sqrt{r_{+}/r_{-}}-\sqrt{r_{-}/r_{+}}\right)\!=
\!{\frac 1 2}\left(\sqrt{\Omega_{-}/\Omega_{+}}-\sqrt{\Omega_{+}/\Omega_{-}}\right)\!=\!\sqrt{M^2/a^2\!-\!1}.
\een
Note that the inverse relation
$r_{\pm}={\sqrt{\Omega_{\mp}/\Omega_{\pm}}}\big/\left(\Omega_{+}+\Omega_{-}\right)$
permits us to replace $r_\pm$ with $\Omega_\pm$ wherever it is necessary,
thus making transparent the duality of the parameters $r_\pm$ and $\Omega_\pm$,
as well as the behavior of the above quantities under interchange of the two horizons:
$r_{+}\leftrightarrows r_{-}$ $\Rightarrow$  $\Omega_{+}\leftrightarrows \Omega_{-}$,
$p\mapsto -p$, $\Omega_{a,g}\mapsto\Omega_{a,g}$ -- invariant.

The parameters $\alpha_{{}_{-}}, \beta_{{}_{-}}, \gamma_{{}_{-}}, \delta_{{}_{-}}, \eta_{{}_{-}}$
can be obtained  by interchanging the places of the two horizons: $r_{+}\!\leftrightarrows r_{-}$
in \eqref{TREparameters:a} -- \eqref{TREparameters:e}.
This procedure may be substantiated using the known properties of the confluent Heun function under
changes of parameters \cite{Heun}. One can check directly that this way we obtain indeed solutions of equation \eqref{TRE}.

According to equations \eqref{TRE16local} and \eqref{TREparameters:f}, the behavior of the solutions
${}_sR^\pm_{\omega,E,m,\sigma_\alpha,\sigma_\beta,\sigma_\gamma}(r;r_{+},r_{-})$
around the corresponding singular points $z=z_{\pm}(r_{\mp};r_{+},r_{-})=0$ is defined by the dominant factor
$\left(z_{{}_\pm}\right)^{\sigma_\beta \beta_{\pm}/2}$.
All other factors in equation \eqref{TRE16local} are regular around these points.
The same solutions
are in general singular around the corresponding singular points $z=z_{\pm}(r_{\pm};r_{+},r_{-})=1$.

Only two of the sixteen solutions  (\ref{TRE16local}) are linearly independent.
Nevertheless, it is necessary to know all of them since for different purposes
one has to use different pairs of independent local solutions.

Using the known asymptotic expansion of the confluent Heun function \cite{Heun}
we obtain two asymptotic solutions of Tom\`{e} type. These are
local solutions of the TRE around its irregular singular point $|r|=\infty$  in the complex plane $\mathbb{C}_r$:
\ben
{}_sR_{\omega,E,m,\sigma_\alpha,\sigma_\beta,\sigma_\gamma}^{\pm\infty}(r;r_{+},r_{-})\sim
{{e^{ i\sigma_{\!\alpha}\,\omega \big(r + (r_{+}+r_{-})\ln r\big)}}}
\sum_{j\geq 0}a_{j}\left(\pm{{r_{+}-r_{-}}\over{r}}\right)^{j+1+(1+\sigma_\alpha)s},\,\,\,\,a_{0}=1.
\la{TMEinf}\een
The notation $\pm\infty$ in \eqref{TMEinf} denotes the two directions: $r\to +\infty$ and  $r\to -\infty$ on the real $r$-axis
for approaching the irregular singular point $|r|=\infty$  in the complex plane $\mathbb{C}_r$.
For the coefficients $a_{j}=a_{j,\omega,E,m,\sigma_\alpha,\sigma_\beta,\sigma_\gamma}$  one has a recurrence relation \cite{Heun}
which shows that they increase together with the integer $j$. Hence, the asymptotic series (\ref{TMEinf}) is a divergent one.

As seen from \eqref{TREparameters:abcdef}, ${}_sR^{-}_{\omega,E,m,\sigma_\alpha,\sigma_\beta,\sigma_\gamma}(r;r_{+},r_{-})=
{}_sR^{+}_{\omega,E,m,\sigma_\alpha,\sigma_\beta,\sigma_\gamma}(r;r_{-},r_{+})$.
Hence, one can introduce a new parity property of the solutions and construct symmetric and anti-symmetric
(with respect to the interchange $r_{+} \rightleftarrows r_{-}$) solutions of the TRE:
\ben
\la{aSymTRE}
{}_sR^{SYM}_{\omega,E,m,\sigma_\alpha,\sigma_\beta,\sigma_\gamma}(r;r_{+},r_{-})={\frac 1 2}
\left({}_sR^{+}_{\omega,E,m,\sigma_\alpha,\sigma_\beta,\sigma_\gamma}(r;r_{+},r_{-})+
{}_sR^{-}_{\omega,E,m,\sigma_\alpha,\sigma_\beta,\sigma_\gamma}(r;r_{+},r_{-})\right),\\
{}_sR^{ASYM}_{\omega,E,m,\sigma_\alpha,\sigma_\beta,\sigma_\gamma}(r;r_{+},r_{-})={\frac 1 2}
\left({}_sR^{+}_{\omega,E,m,\sigma_\alpha,\sigma_\beta,\sigma_\gamma}(r;r_{+},r_{-})-
{}_sR^{-}_{\omega,E,m,\sigma_\alpha,\sigma_\beta,\sigma_\gamma}(r;r_{+},r_{-})\right).
\nonumber
\een
Clearly, these solutions are singular at both horizons in the general case.
When one considers the two-singular-point boundary problem \cite{Heun}
on the interval $[r_{-},r_{+}]$  in the Kerr black hole interior,
the solutions \eqref{aSymTRE} may be regular at one, or at both the ends
for some values of the separation constants $\omega$ and $E$.
Since this boundary problem is still not studied,
at present we are not able to make more definite statements about this case.

\section{A New Classification of the Solutions to the TRE,
Based on the $\delta_N$-Condition. Novel Radial $\delta_N$-Solutions}

For the TRE  the $\delta_N$-condition reads:
\ben
{}_s\omega^\pm_{m,\sigma_\alpha,\sigma_\beta,\sigma_\gamma}\, {\cal L}^{\pm}_{\sigma_\alpha,\sigma_\beta,\sigma_\gamma}=
\Omega_g\left({\cal M}^{\pm}_{m,\sigma_\alpha,\sigma_\beta,\sigma_\gamma} +
i\, {}_s{\cal N}^\pm_{\sigma_\alpha,\sigma_\beta,\sigma_\gamma}\right),
\la{TRE_delta_cond}
\een
where $${\cal L}^{\pm}_{\sigma_\alpha,\sigma_\beta,\sigma_\gamma}\!=\!
{\frac{\sigma_\beta \Omega_{\pm}\!-\!\sigma_\gamma\Omega_{\mp}}{\Omega_{\pm}\!-\!\Omega_{\mp}}}\!-\!\sigma_\alpha,\,\,
{\cal M}^{\pm}_{m,\sigma_\alpha,\sigma_\beta,\sigma_\gamma}\!=\!m(\sigma_\beta\!-\!\sigma_\gamma){\frac{\Omega_g}{\Omega_{\pm}\!-\!\Omega_{\mp}}},\,\,
{}_s{\cal N}^\pm_{\sigma_\alpha,\sigma_\beta,\sigma_\gamma}\!=\!N\!+\!1\!+\!\left(\sigma_\alpha\!+\!{\frac{\sigma_\beta\!+\!\sigma_\gamma} 2}\right)s.$$

We call {\em radial $\delta_N$-solutions} the solutions defined  via the $\delta_N$-condition \eqref{TRE_delta_cond}.

The calculation of the values of the coefficients in equation (\ref{TRE_delta_cond})
yields two very different cases:

1. In the first case  ${\cal L}^{+}_{\pm,\pm,\pm}={\cal L}^{-}_{\pm,\pm,\pm}=0$
and we see that one is not able to fix the frequencies
${}_s\omega^{+}_{m,\pm,\pm,\pm}$ and ${}_s\omega^{-}_{m,\pm,\pm,\pm}$.
Instead, choosing $\sigma_\alpha=\sigma_\beta = \sigma_\gamma =-\sigma$
and using (\ref{TRE_delta_cond}) one fixes the non-negative integer $N$ in the form
\ben
{}_sN+1= 2|s|\geq 1 \,\,\,\text{for}\,\,\,|s|\geq 1/2.
\la{sN_Kerr}
\een

Thus, the degree of the polynomial $\Delta_{N+1}$-condition is fixed, too.

2. In the second case the coefficients ${\cal L}^{\pm}_{\sigma_\alpha,\sigma_\beta,\sigma_\gamma}$ are nonzero and
one can fix the values of the frequencies ${}_s\omega^\pm_{m,\sigma_\alpha,\sigma_\beta,\sigma_\gamma}$
from equation (\ref{TRE_delta_cond}). Thus, one obtains two different types of exact {\em equidistant} spectra:

a) For ${\cal L}^{+}_{\mp,\pm,\pm}={\cal L}^{-}_{\mp,\pm,\pm}=\pm 2$, ${\cal M}^{+}_{\mp,\pm,\pm}={\cal M}^{-}_{\mp,\pm,\pm}=0$ and
${\cal N}^{+}_{\mp,\pm,\pm}={\cal N}^{-}_{\mp,\pm,\pm}=(N+1)$ the
\vskip.03truecm
\noindent $\delta_N$-condition (\ref{TRE_delta_cond})
produces the pure imaginary equidistant frequencies with $N\geq 0$ -- integer:
\ben
{}_s\omega^{+}_{N,m,\mp,\pm,\pm}={}_s\omega^{-}_{N,m,\mp,\pm,\pm}=\pm i\,{ (N+1) /{4M}}=\pm i\,\Omega_g (N+1)/2.
\la{Im_omega_N_Kerr}
\een
Note that these frequencies depend neither on the spin-weight  $s$ and azimuthal number $m$, nor on the rotation parameter $a$.
The spectrum is not influenced  by the rotation of the waves and of the very Kerr metric.
The frequencies (\ref{Im_omega_N_Kerr}) are defined only by the monopole term in multipole expansion of the metric.

b) For all other cases the coefficients ${\cal L}^{\pm}_{\sigma_\alpha,\sigma_\beta,\sigma_\gamma}$,                                                       and ${\cal M}^{\pm}_{\sigma_\alpha,\sigma_\beta,\sigma_\gamma}$
are not fixed and one obtains the following two similar double-equidistant spectra of frequencies
with $N\geq 0\text{ -- integer}$ and $m\,\text{ -- (half)integer}$:
\begin{subequations}\label{omega_N:ab}
\ben
{}_s\omega^{+}_{N,m,\mp,\mp,\pm}={}_s\omega^{-}_{N,m,\mp,\pm,\mp}
=\Omega_{+}\left(m \pm i p (N+1 \mp s)\right);
\label{omega_N:a}\\
{}_s\omega^{+}_{N,m,\pm,\mp,\pm}={}_s\omega^{-}_{N,m,\pm,\pm,\mp}
=\Omega_{-}\left(m \pm i p (N+1 \pm s)\right). \label{omega_N:b}
\een
\end{subequations}

A set of important new mathematical properties of the radial $\delta_N$-solutions can be found in \cite{Fiziev2009a,Fiziev2009b}.
In \cite{Fiziev2009b} it is shown that these solutions define the most general class
of solutions to the TRE for which the properly generalized Teukolsky-Starobinsky identities exist.
The solutions which satisfy the relation \eqref{sN_Kerr} were studied
in \cite{Teukolsky, Chandra} without utilizing the Heun functions and the $\delta_N$-condition.
The last condition turns to be valid automatically for the solutions to the TRE studied in \cite{Teukolsky, Chandra}.
The infinite series of the solutions with equidistant spectra \eqref{Im_omega_N_Kerr} and \eqref{omega_N:ab}
are introduced and considered for the first time in the present paper.
In the third of the papers \cite{PFDS_astroph:HE} one can find an interesting
and unexpected recent application of the formulas \eqref{omega_N:ab}  for fitting
of the spectra of electromagnetic jets from Kerr black holes and necked singularities.

\section{Polynomial Solutions to the TRE}

The $\delta_N$-condition yields the basic classification of the solutions described in the previous section 3.
As a result, one obtains two classes of polynomial solutions to the TRE,
imposing in addition the $\Delta_{N+1}$-condition \eqref{PolynomCond:b}.
In what follows we will use the determinant form of the $\Delta_{N+1}$-condition given in \cite{Fiziev2009a}.

\subsection{The First Class of Polynomial Solutions to the TRE:}

The solutions of this class correspond to the first case in section 3 and obey
equation (\ref{sN_Kerr}).
The inequality  ${}_sN=2|s|-1 \geq 0$ excludes the existence of scalar perturbations ($|s|=0$)
of the first polynomial class.

\subsubsection{The General Case:}
For brevity, we denote the solutions ${}_sR^{\pm}_{\omega,E,m,-\sigma,-\sigma,-\sigma}(r;r_{+},r_{-})$
as ${}_sR^{\pm}_{\omega,E,m}(r;r_{+},r_{-})$.
For them the parameter $\mu$ takes the values
$\mu={}_s\mu_{\omega,k,m}^{\pm}(r_{+},r_{-}),\,\,\, k=1,\dots, 2|s|$ --
the solutions of the algebraic equation \eqref{PolynomCond:b}, which now takes the form: $\Delta^{\pm}_{2|s|}(\mu)=0$.
Its degree is $2|s|=1, 2, 3$, or $4$, depending on the spin of the perturbations $|s|=1/2, 1, 3/2, 2$.
Making use of \eqref{munu:b}, and \eqref{TREparameters:a}-\eqref{TREparameters:e},
we obtain for the separation constant $E={}_sE_{\omega,k,m}^{\pm}(r_{+},r_{-})$,
$k\!=\!1,\dots, 2|s|$ the expressions
\ben
{}_sE_{\omega,k,m}^{\pm}(r_{+},r_{-})\!=\!{}_s\mu_{\omega,k,m}^{\pm}(r_{+},r_{-})
+|s|(|s|-1)-a\omega(a\omega-2m)+2i\sigma(2|s|-1)\omega r_{\mp}, \label{E_first_class}
\een

Applying the explicit expressions for the roots ${}_s\mu_{\omega,k,m}^{\pm}(r_{+},r_{-})$,
we obtain:
\ben
{}_sE_{\omega,m}^{\pm}(r_{+},r_{-})\!=-a^2\omega^2+2a\omega m-{\frac 1 4}:\,\,\,\text{for}\,\,\,|s|={\frac 1 2},\, m=\pm 1/2, \pm 3/2,\dots;
\la{E_first_class_1/2}
\een
\ben
{}_sE_{\omega,k,m}^{\pm}(r_{+},r_{-})\!=-a^2\omega^2+2a\omega \left(m-(-1)^k \sqrt{1-m/a\omega}\right)\!:
\,\,\text{for}\,\,\,k=1,2,\,|s|=1,\,\,m=\pm 1, \pm 2, \dots
\la{E_first_class_1}
\een

For the gravitational waves ($|s|=2$) one has  to find the quantities
${}_s\mu_{\omega,k,m}^{\pm}(r_{+},r_{-})$ solving algebraic equation
of the fourth degree $\Delta^{\pm}_{4}(\mu)=0$.
The explicit form of its roots is too complicated and not necessary for the purposes of the present paper.
It is more instructive to demonstrate here the result,
obtained using the Taylor series expansion of the solutions ${}_s\mu_{\omega,k,m}^{\pm}(r_{+},r_{-})$
around the zero frequency $\omega=0$.

Thus, we obtain  for $|s|=2,\,\,k=1,2$, and $m=\pm 2, \pm 3, \dots$ the eight series of values:
\ben\label{mu:1_2}
{}_sE_{\omega,k,m}^{\pm}\!=\!2\!-\!4\left(m\!-\!i(-1)^k {\frac{3M}{2a}}\right)a\omega\!+\!
6\!\left(\!m^2\!+\!i(-1)^k2 m\left((m^2\!-\!1){\frac a M}\!+\!{\frac {2M} {a}}\right)\!+\!{\frac{3M^2}{a^2}}\!-\!{\frac{7}{6}}\right)(a\omega)^2
\!+\\
+{\cal{O}}_{3}(a\omega).\nonumber
\een
For $|s|=2,\,\,k=3,4$, and $m=\pm 2, \pm 3, \dots$ we have other eight series of values:
\ben\label{mu:3_4}
{}_sE_{\omega,k,m}^{\pm}=
i\,(-1)^k 4 \sqrt{ma\omega}\Bigg(1+i\,3\left(1+\left({\frac{3M^2}{8a^2}}-{\frac{2}{3}}\right){\frac{1}{m^2}}\right)ma\omega
+\!{\cal{O}}_{2}(a\omega)\Bigg)+\hskip 2.7truecm\nonumber\\
+8 ma \omega-6\left(1+\left({\frac{3M^2}{a^2}}-{\frac{5}{6}}\right){\frac{1}{m^2}}\right)(ma\omega)^2+\!{\cal{O}}_{3}(a\omega).
\hskip .5truecm
\een

Clearly, these series describe two kinds of solutions with a completely different behavior around the origin $\omega=0$.
In particular, the series \eqref{mu:1_2} and \eqref{mu:3_4} have different limits: $2$ and $0$, respectively, when $\omega\to 0$.
For the solutions \eqref{mu:3_4} the origin $\omega=0$ is a branching point, etc.

The independence of the values of ${}_sE_{\omega,k,m}^{\pm}$ in \eqref{mu:1_2} and \eqref{mu:3_4} on the upper labels $(\pm)$
is a result of the polynomial character of the solutions, i.e. of the regularity of the corresponding HeunC-factor
simultaneously on both the horizons $r_{\pm}$.

For a complete solution of the problem one has to determine the frequency $\omega$. Hence,
one needs an additional relation between the parameters $E$ and $\omega$.
This relation may appear when one solves the TAE (See the next sections 5-8.).

The first class of polynomial solutions to the TRE is introduced and studied in detail for the first time in the present
paper.

\subsubsection{The Special Case of the Schwarzschild metric:}
For the special value of the parameter  $a=0$ we have $r_{-}=0$, $r_{+}=2M$.
This is the case of perturbations to the nonrotating Schwarzschild black hole described in terms of the Weyl scalars.
For simplicity, here we use units in which $2M=1$.
The parameters in the solution (\ref{TRE16local}) acquire the limiting values
\ben
\alpha_{+}=2i\omega, \beta_{+}=s, \gamma_{+}=s+2i\omega, \delta_{+}=2i\omega(s-i\omega), \eta_{+}=-E+{\frac{s^2}{2}};
\hskip 1.3truecm \nonumber\\
\alpha_{-}=-2i\omega, \beta_{-}=s+2i\omega, \gamma_{-}=s, \delta_{-}=-2i\omega(s-i\omega),
\eta_{-}=-E+{\frac{s^2}{2}}+2\omega^2+2is\omega.
\la{parameters_Schwarcschild}
\een
These differ from the values of the parameters of confluent Heun's functions
in the Regge-Wheeler approach to the perturbations of the Schwarzschild metric \cite{F}.

In the limit $a\to 0$ equation \eqref{TRE_delta_cond} does not define the frequency $\omega$,
if $\sigma_\alpha=\mp\sigma_\beta=\pm\sigma_\gamma=-\sigma$,
because then one obtains ${\cal L}^{\pm}_{-\sigma,\pm\sigma,\mp\sigma}=0$.
If, in addition, $\sigma=\text{sign}(s)$, then the $\delta_N$-condition is fulfilled for the special
polynomial solutions of the first class denoted as
${}_sR^{\pm}_{\omega,E,m}(r)={}_sR^{\pm}_{\omega,E,m,-\sigma,\pm\sigma,\mp\sigma}(r;1,0)$.
Equation \eqref{TRE_delta_cond} yields the relation ${}_sN=|s|-1 \geq 0$.
Scalar perturbations of this type do not exist.

In the case of integer spins $|s|=1,2$ the roots $\mu={}_s\mu_{\omega,k,m}^{\pm},\,\,\, k=1,\dots, |s|$
of the equations $\Delta^{\pm}_{|s|}(\mu)=0$,  \eqref{munu:b},
and \eqref{TREparameters:a}-(\ref{TREparameters:e}) with $r_{+}=1$, $r_{-}=0$ and $a=0$
produce the following simple expressions for $E={}_sE_{\omega,k,m}^{\pm}$, $k=1,\dots, |s|$:
\begin{eqnarray}
{}_sE_{\omega,m}^{\pm}&=&0:\,\,\,\text{for}\,\,\, |s|=1,\\
{}_sE_{\omega,k,m}^{\pm}&=&1-(-1)^k\sqrt{1-i6\sigma\omega}:\,\,\,\text{for}\,\,\, |s|=2,\,k=1,2.
\la{A_first_special}
\end{eqnarray}
For a complete solution of the problem, one needs an additional relation between the parameters $E$ and $\omega$.
This relation may be found by solving the TAE, see sections 5-8.

The above considerations of the limit  $a\to 0$ and the corresponding results for the Schwarzschild black hole
in terms of confluent Heun's functions are new
and obtained for the first time in the present paper.

\subsection{Second Class of Polynomial Solutions to the TRE:}
According to the results of section 3, the solutions of this class originate from the second case
of the $\delta_N$-condition and fall into two subclasses: a) and b).
The complete definite frequencies
${}_s\omega^{\pm}_{N,m,\sigma_\alpha,\sigma_\beta,\sigma_\gamma}$
-- formulae \eqref{Im_omega_N_Kerr} and \eqref{omega_N:ab},
yield algebraic equations $\Delta^{\pm}_{N+1}(\mu)=0$ with
$(N+1)$ roots $\mu={}_s\mu^{\pm}_{N,n,m,\sigma_\alpha,\sigma_\beta,\sigma_\gamma}(r_{+},r_{-})$, $n=0, 1, \dots, N$.
It seems difficult to derive explicit analytic expressions for these roots, but their numerical values can be easily obtained.
Using the values of ${}_s\mu^{\pm}_{N,n,m,\sigma_\alpha,\sigma_\beta,\sigma_\gamma}(r_{+},r_{-})$
and equations \eqref{munu:b}, \eqref{TREparameters:a}-\eqref{TREparameters:e}
we obtain complete {\em definite} values for the parameter
$E={}_sE_{N,n,m,\sigma_\alpha,\sigma_\beta,\sigma_\gamma}^{\pm}(r_{+},r_{-})$:

a) In the case of frequencies \eqref{Im_omega_N_Kerr} we obtain
\ben\label{1E_second_class}
{}_sE_{N,n,m,\sigma,-\sigma,-\sigma}^{\pm}\!=\!{}_s\mu_{N,n,m,\sigma,-\sigma,-\sigma}^{\pm}+
|s|(|s|-1)+a\omega(3a\omega-2m)\!+4\,\omega^2r_{\mp}^2\!+\!2i\sigma\omega\big(2M|s|-r_{\pm}\big).
\een

b) In the case of frequencies  \eqref{omega_N:a}, \eqref{omega_N:b} we have, respectively:
\begin{subequations}\label{2E_second_class:ab}
\ben
{}_sE_{N,n,m,+,-,+}^{\pm}\!-{}_s\mu_{N,n,m,+,-,+}^{\pm}={}_sE_{N,n,m,-,+,-}^{\mp}\!-{}_s\mu_{N,n,m,+,-,+}^{\mp}=
\pm i\,{2(2a\omega-m)}/{2 p}-\hskip 1.95truecm\nonumber \\
-\Big(m^2+8m\left(1+M^2/a^2\right)a\omega+
\left(1+ 10\, r_{\mp}/r_{\pm}+9\,(r_{\mp}/r_{\pm})^2-4\,(r_{\mp}/r_{\pm})^3\right)\omega^2 r_{\pm}^2\Big)\big/{4 p^2},
\hskip 1.truecm\label{2E_second_class.a}\\
{}_sE_{N,n,m,+,+,-}^{\pm}\!-\!{}_s\mu_{N,n,m,+,+,-}^{\pm}\!=\!{}_sE_{N,n,m,-,-,+}^{\mp}\!-\!{}_s\mu_{N,n,m,-,-,+}^{\mp}\!=\!
\pm i\,2\Big(m\!+\!2a\omega\!\left(1\!-\!2M^2/a^2\right)\!\Big)\big/2 p-\hskip .2truecm\nonumber\\
-i2psa\omega-4\Big(m^2+2m\left(1\!-\!3M^2/a^2\right)a\omega-\left(1\!-\!5M^2/a^2\right)(a\omega)^2\Big)\big/4 p^2.
\hskip 1.truecm\label{2E_second_class.b}
\een
\end{subequations}

With $\omega$ and $E$ given by equations \eqref{Im_omega_N_Kerr},
\eqref{omega_N:ab} and \eqref{1E_second_class}, \eqref{2E_second_class:ab}
we have no more free parameters in the problem at hand.
As a result, the corresponding solutions to the TAE are fixed unambiguously
by the designated group of equations
obtained for the second class of polynomial solutions to the TRE.
This situation is completely new,
unexpected and described here for the first time.

\section{Exact Solutions to the Teukolsky Angular Equation in Terms of the Confluent Heun functions}

In terms of the variable $x=\cos\theta$ the TAE has three singular points.
Two of them: $x_{-}=-1$ (i.e., $\theta_{S}=\pi$ -- South (S-)pole)
and $x_{+}= 1$ (i.e., $\theta_{N}=0$ -- North (N-)pole) are regular singular points.
The third one $x_\infty=\infty$ is an irregular singular point.
It is remarkable that introducing the notation
\ben
z_{+}=z_{+}(\theta)=\left(\cos(\theta/2)\right)^2,\,\,\, z_{-}=z_{-}(\theta)=\left(\sin(\theta/2)\right)^2, \,\,\,z_{+}+z_{-}=1;
\la{TAE_zpm}
\een
and
\ben
a_{\pm}=\pm 4a\omega,\,\,b_{\pm}=s\mp m,\,\,c_{\pm}=s\pm m,\,\,d_{\pm}=\pm 4s a\omega ,\,\,n_{\pm}={\frac{m^2+s^2}{2}}\mp 2s a\omega -a^2\omega^2-E.
\label{TAEparameters}
\een
we can write down 16 local solutions of the TAE in the form
\ben
{}_sS_{\omega,E,m,\sigma_a,\sigma_b,\sigma_c}^{\pm}=
\!e^{\sigma_a{\frac{a_{{}_\pm} z_{{}_\pm}} 2}}
z_{{}_\pm}^{\sigma_b{\frac{b_{{}_\pm}} 2}} z_{{}_\mp}^{\sigma_c{\frac{c_{{}_\pm}} 2}}
\text{HeunC}(\sigma_a a_{{}_\pm},\sigma_b b_{{}_\pm},\sigma_c c_{{}_\pm},d_{{}_\pm},n_{{}_\pm},z_{{}_\pm})
\la{TAE16local}
\een
which is very similar to the form of Eq. \eqref{TRE16local}.

Following the corresponding properties of the TAE \eqref{angE:a} \cite{Teukolsky}, the solutions \eqref{TAE16local} have the symmetries
\begin{subequations}\label{symmetriesTAE16local:ab}
\ben
{}_{-s}S_{\omega,E,m,\sigma_a,\sigma_b,\sigma_c}^{\pm}(\pi-\theta)&=&
{}_sS_{\omega,E,m,-\sigma_a,-\sigma_b,-\sigma_c}^{\mp}(\theta),
\label{symmetriesTAE16local:a}\\
{}_{s}S_{-\omega,E,-m,\sigma_a,\sigma_b,\sigma_c}^{\pm}(\pi-\theta)&=&
{}_sS_{\omega,E,m,\sigma_a,\sigma_b,\sigma_c}^{\mp}(\theta).
\label{symmetriesTAE16local:b}
\een
\end{subequations}

Note that according to Eq. \eqref{TAE16local}, the behavior of the solutions
${}_sS_{\omega,E,m,\sigma_a,\sigma_b,\sigma_c}^{\pm}$
around the corresponding singular points $z=z_{+}(\theta_{S})=z_{-}(\theta_{N})=0$ is defined by the dominant factor
$\left(z_{{}_\pm}\right)^{\sigma_b b_{{}_\pm}/2}$.
All other factors in \eqref{TAE16local} are regular around these points.
The same solutions
are in general singular around the corresponding singular points $z=z_{+}(\theta_{N})=z_{-}(\theta_{S})=1$.
Hence, at this point we have a complete analogy with the case of the TRE.

Only two of the sixteen solutions  \eqref{TAE16local} are linearly independent.
Nevertheless, it is important to know all of them, since for various purposes
one can use different pairs of independent local solutions, see below.
If one chooses some two linearly independent
solutions, then one can represent the other fourteen using this basis. Unfortunately,
at present the form of the corresponding coefficients is completely unknown.

We can establish simple relations between some of the different solutions \eqref{TAE16local}
in proper domains of the parameters $s$ and $m$, if we divide the whole plane $\{s,m\}$ into four sectors.
In each of them we choose the solutions with the same {\em regular} asymptotic behavior around the corresponding pole
as follows:

I. Sector $s\geq 0$, $|m|\leq |s|$:
\begin{subequations}\label{sectorI:ab}
\ben
{}_sS^{+\, reg}_{\omega,E,m}(\theta)\!&\!=\!&\!{}_sS^{+}_{\omega,E,m,+++}\!=\!{}_sS^{+}_{\omega,E,m,-++}\!=\!
{}_sS^{+}_{\omega,E,m,++-}\!=\!
{}_sS^{+}_{\omega,E,m,-+-}\,\underset{\theta\to\pi}{\sim}\left(\cos{\frac\theta 2}\right)^{s-m},\label{sectorI:a}\\
{}_sS^{-\, reg}_{\omega,E,m}(\theta)\!&\!=\!&\!{}_sS^{-}_{\omega,E,m,+++}\!=\!{}_sS^{-}_{\omega,E,m,-++}\!=\!
{}_sS^{-}_{\omega,E,m,++-}\!=\!
{}_sS^{-}_{\omega,E,m,-+-}\,\underset{\theta\to 0}{\sim}\left(\sin{\frac\theta 2}\right)^{s+m}.\label{sectorI:b}
\een
\end{subequations}

II. Sector $m\leq 0$, $|s|\leq |m|$:
\begin{subequations}\label{sectorII:ab}
\ben
{}_sS^{+\, reg}_{\omega,E,m}(\theta)\!&\!=\!&\!{}_sS^{+}_{\omega,E,m,+++}\!=\!{}_sS^{+}_{\omega,E,m,-++}\!=\!
{}_sS^{+}_{\omega,E,m,++-}\!=\!
{}_sS^{+}_{\omega,E,m,-+-}\,\underset{\theta\to\pi}{\sim}\left(\cos{\frac\theta 2}\right)^{s-m},\label{sectorII:a}\\
{}_sS^{-\, reg}_{\omega,E,m}(\theta)\!&\!=\!&\!{}_sS^{-}_{\omega,E,m,---}\!=\!{}_sS^{-}_{\omega,E,m,+--}\!=\!
{}_sS^{-}_{\omega,E,m,--+}\!=\!
{}_sS^{-}_{\omega,E,m,+-+}\,\underset{\theta\to 0}{\sim}\left(\sin{\frac\theta 2}\right)^{-s-m}.\label{sectorII:b}
\een
\end{subequations}

III. Sector $s\leq 0$, $|m|\leq |s|$:
\begin{subequations}\label{sectorIII:ab}
\ben
{}_sS^{+\, reg}_{\omega,E,m}(\theta)\!&\!=\!&\!{}_sS^{+}_{\omega,E,m,---}\!=\!{}_sS^{+}_{\omega,E,m,+--}\!=\!
{}_sS^{+}_{\omega,E,m,--+}\!=\!
{}_sS^{+}_{\omega,E,m,+-+}\,\underset{\theta\to\pi}{\sim}\left(\cos{\frac\theta 2}\right)^{-s+m},\label{sectorIII:a}\\
{}_sS^{-\, reg}_{\omega,E,m}(\theta)\!&\!=\!&\!{}_sS^{-}_{\omega,E,m,---}\!=\!{}_sS^{-}_{\omega,E,m,+--}\!=\!
{}_sS^{-}_{\omega,E,m,--+}\!=\!
{}_sS^{-}_{\omega,E,m,+-+}\,\underset{\theta\to 0}{\sim}\left(\sin{\frac\theta 2}\right)^{-s-m}.\label{sectorIII:b}
\een
\end{subequations}

IV. Sector $m\geq 0$, $|s|\leq |m|$:
\begin{subequations}\label{sectorIV:ab}
\ben
{}_sS^{+\, reg}_{\omega,E,m}(\theta)\!&\!=\!&\!{}_sS^{+}_{\omega,E,m,---}\!=\!{}_sS^{+}_{\omega,E,m,+--}\!=\!
{}_sS^{+}_{\omega,E,m,--+}\!=\!
{}_sS^{+}_{\omega,E,m,+-+}\,\underset{\theta\to\pi}{\sim}\left(\cos{\frac\theta 2}\right)^{-s+m},\label{sectorIV:a}\\
{}_sS^{-\, reg}_{\omega,E,m}(\theta)\!&\!=\!&\!{}_sS^{-}_{\omega,E,m,+++}\!=\!{}_sS^{-}_{\omega,E,m,-++}\!=\!
{}_sS^{-}_{\omega,E,m,++-}\!=\!
{}_sS^{-}_{\omega,E,m,-+-}\,\underset{\theta\to 0}{\sim}\left(\sin{\frac\theta 2}\right)^{s+m}.\label{sectorIV:b}
\een
\end{subequations}

Note that in each sector the four solutions in the above relations of type (a),
or in the above relations of type (b) are equal,
since under standard normalization the local {\em regular} solution
around any regular singular point of the TAE is unique.

In the case of the TAE there exist an additional complication.
The numbers $s$ and $m$ are {\em simultaneously} integers, or half-integers.
Then $\beta=\sigma_b b_{{}_\pm}=\sigma_b (s\mp m)$ is an integer and, in particular, it {\em may} be a negative integer.
However, the confluent Heun functions $\text{HeunC}(\alpha,\beta,\gamma,\delta,\eta,z)$
are not defined when $\beta$ is a negative integer \cite{Heun}.
Therefore, if $\beta=\sigma_b b_{{}_\pm}<0$ is a negative integer,
we must write down the corresponding solutions in the form
\ben
{}_sS_{\omega,E,m,\sigma_a,\sigma_b,\sigma_c}^{\pm}=
\!e^{\sigma_a{\frac{a_{{}_\pm} z_{{}_\pm}} 2}}
z_{{}_\pm}^{\sigma_b{\frac{b_{{}_\pm}} 2}} z_{{}_\mp}^{\sigma_c{\frac{c_{{}_\pm}} 2}}
\underline{\text{HeunC}}(\sigma_a a_{{}_\pm},\sigma_b b_{{}_\pm},\sigma_c c_{{}_\pm},d_{{}_\pm},n_{{}_\pm},z_{{}_\pm}).
\la{TAE16local_cc}
\een
For this purpose we define the {\em concomitant} confluent Heun function\footnote{Note that for
any value of the parameter $\beta$, when the confluent Heun function in the right hand side of Eq. \eqref{HeunCcc}
is well defined, its left hand side represents a second, linearly independent solution
of the confluent Heun equation.}
\ben
\underline{\text{HeunC}}(\alpha,\beta,\gamma,\delta,\eta,z)=z^{-\beta}\text{HeunC}(\alpha,-\beta,\gamma,\delta,\eta,z)
\int{\frac{e^{-\alpha\zeta}\zeta^{\beta-1}(1-\zeta)^{-\gamma-1}}{\big(\text{HeunC}(\alpha,-\beta,\gamma,\delta,\eta,\zeta)\big)^2}}d\zeta.
\la{HeunCcc}
\een
This function is well defined for negative integer $\beta=\sigma_b b_{{}_\pm}< 0$,
together with the confluent function $\text{HeunC}(\alpha,-\beta,\gamma,\delta,\eta,z)$.
In this case, the function $z^{-\beta}\text{HeunC}(\alpha,-\beta,\gamma,\delta,\eta,z)$
represents the local regular solution to the confluent Heun equation \eqref{DHeunC} around the singular point $z=0$
and the concomitant confluent function  $\underline{\text{HeunC}}(\alpha,\beta,\gamma,\delta,\eta,z)$
represents a second linearly independent local solution, which is singular around this point. We need the concomitant
Heun function to construct second independent local solution in the case $\beta=0$, too, since in this case
$z^{-\beta}\text{HeunC}(\alpha,-\beta,\gamma,\delta,\eta,z)\equiv \text{HeunC}(\alpha,\beta,\gamma,\delta,\eta,z)$
can not be used for this purpose.

It can be shown that for negative integer $\beta$ the concomitant confluent Heun function has the form
\ben
\underline{\text{HeunC}}(\alpha,\beta,\gamma,\delta,\eta,z)=\sum_{n=1}^{|\beta|}{\frac{c_n}{z^n}} + h_1(z)+h_2(z)\ln(z),
\,\,\,{\text{all}}\,\,\,c_n\neq 0.
\la{ccH}
\een
Here $h_{1,2}(z)$ denote two definite functions of the complex variable $z$
which are analytic in the vicinity of the point $z=0$.
In the problem at hand $|\beta|=|\beta_\pm|=|s\mp m|$.
The logarithmic term is present in the concomitant confluent Heun function when $|\beta|=0$, too,
but then we have no poles in the solution \eqref{ccH}. For $|\beta|=0$ its form otherwise is similar to (\ref{ccH}).
One can reach the last results using general analytical methods described, for example, in \cite{DE}.

The above consideration sows that in the case of the TAE we can construct only eight local solutions \eqref{TAE16local}
which are single-valued functions of the variable $z$. The other eight solutions, being in the form \eqref{TAE16local_cc},
are infinitely-valued, because of the logarithmic term in Eq. \eqref{ccH}.

\section{A New Classification of the Solutions to the TAE Based on the  $\delta_N$-Condition.
Novel $\delta_N$-Angular Solutions}

For solutions ${}_sS_{\omega,E,m,\sigma_a,\sigma_b,\sigma_c}^{\pm}$ (\ref{TAE16local}) to the TAE  the $\delta_N$-condition reads:
\ben
0=\mp\, m{\frac{\sigma_b-\sigma_c}{2}} + N+1+\left(\sigma_a + {\frac{\sigma_b+\sigma_c}{2}} \right)s.
\la{TAE_delta_cond}
\een

We call {\em angular $\delta_N$-solutions} the solutions
defined via the $\delta_N$-condition \eqref{TAE_delta_cond}.
To some extent these solutions are similar to the radial $\delta_N$-solutions introduced in section 3.
A set of important new mathematical properties of the angular $\delta_N$-solutions
can be found in \cite{Fiziev2009a,Fiziev2009b}.
In \cite{Fiziev2009b} it is shown that these solutions define the most general class
of solutions to the TAE for which properly generalized Teukolsky-Starobinsky identities exist.

Comparing equation \eqref{TAE_delta_cond} with the corresponding one for the TRE
-- \eqref{TRE_delta_cond}, we see both essential differences and similarities.
For the coefficients in equation \eqref{TAE_delta_cond}, which are analogous to the ones in \eqref{TRE_delta_cond},
one obtains:
$${\cal L}^{\pm}_{\sigma_a,\sigma_b,\sigma_c}\equiv 0,\,\,
{\cal M}^{\pm}_{m,\sigma_a,\sigma_b,\sigma_c}=\mp \,m(\sigma_b\!-\!\sigma_c){\frac 1 2},\,\,
{}_s{\cal N}^\pm_{\sigma_a,\sigma_b,\sigma_c}=N\!+\!1\!+\!\left(\sigma_a\!+\!{\frac{\sigma_b\!+\!\sigma_c} 2}\right)s.$$

Hence:

i) The coefficients ${\cal L}^{\pm}_{\sigma_a,\sigma_b,\sigma_c}$ vanish identically,
in contrast to the coefficients  ${\cal L}^{\pm}$ in equation \eqref{TRE_delta_cond}.
Consequently, there are no cases in which the condition \eqref{TAE_delta_cond} can fix the frequencies $\omega$.

ii) The form of the coefficients ${\cal M}^{\pm}$ of both equations \eqref{TRE_delta_cond} and \eqref{TAE_delta_cond}
is the same only for $M/a=\sqrt{2}$.

iii) The coefficients ${\cal N}^\pm$ of both equations are of the same form.

We obtain two different cases depending on the coefficient $(\sigma_b\!-\!\sigma_c)$ in front of the azimuthal number $m$:

1. The first class angular $\delta_N$-solutions with $\sigma_c=\sigma_b$ and
$\sigma_a=\sigma_b=\sigma_c=-\sigma$. As a result,
Eq. \eqref{TAE_delta_cond} fixes the degree of the second polynomial
condition $\Delta_{N+1}=0$ in the same form as equation \eqref{sN_Kerr}\footnote{The alternative case
$\sigma_b=\sigma_c=-\sigma_a$ leads to a uninteresting relation $N+1=0$.}:
\ben
{}_sN+1= 2|s|\geq 1\,\,\,\text{for}\,\,\,|s|\geq 1/2.
\la{sN_Kerr_A}
\een

In the case of the TAE the set of $\delta_N$-solutions  consists of
ones with {\em integer} parameters $\beta =\sigma_b b_{{}_\pm}$ of both signs.
According to Eqs. \eqref{TAE16local_cc} and \eqref{ccH}, for negative integer $\sigma_b b_{{}_\pm} < 0$ in the solutions we have
logarithmic terms. Such solutions are infinitely-valued functions. To exclude this physically
not admissible case, one must impose the additional requirement $\sigma_b b_{{}_\pm} \geq 0$.
As a result one obtains $\sigma\sigma_m=\pm 1$, $\sigma_b b_{{}_\pm}=|m|-|s|\geq 0$ and $\sigma_c c_{{}_\pm}=-|m|-|s|<0$.
Thus, the single-valued angular $\delta_N$-solutions of the first class with spin $|s|\geq 1/2$
correspond to sectors II \eqref{sectorII:ab} and IV \eqref{sectorIV:ab} and acquire the form
\ben
\la{TAE_delta_N_integer_S_A}
{}_sS_{\omega,E,m}^{\pm}(z_{{}_\pm})=
\!e^{-2\sigma_m a \omega z_{{}_\pm} }
\left(z_{{}_\pm}\right)^{\frac{|m|-|s|}2} \left(z_{{}_\mp}\right)^{\frac{-|m|-|s|}2}\times \hskip 6.4truecm\\
\times\text{HeunC}(-4\sigma_m a \omega,|m|-|s|,-|m|-|s|, 4\sigma_m a\omega|s|,{\frac{m^2+s^2}{2}}- 2\sigma_m a\omega|s| -a^2\omega^2-E,z_{{}_\pm}).
\nonumber
\een

Obviously, the solutions \eqref{TAE_delta_N_integer_S_A}
are regular around the singular points $z_{{}_\pm}\!=\!0$.
Their behavior around the singular points $z_{{}_\pm}\!=\!1 (\Leftrightarrow z_{{}_\mp}=0$) is more complicated.
We can study this behavior using the expansion of the solutions \eqref{TAE_delta_N_integer_S_A}
with respect to the  basis of the two linearly independent local solutions \eqref{TAE16local_cc}
around the points $z_{{}_\mp}\!=\!0$, which are well defined for $\sigma\sigma_m=\pm 1$,
$\sigma_b b_{{}_\pm}\!=\!|m|-|s|\geq 0$ and $\sigma_c c_{{}_\pm}\!=\!-|m|-|s|<0$:
\ben
\la{TAE_delta_N_integer_S_expansion}
{}_sS_{\omega,E,m}^{\pm}(z_{{}_\pm})=
{}_s\Gamma^{\pm}_1(\omega,E,m)\, e^{ 2\sigma_m a \omega z_{{}_\mp}}
\left(z_{{}_\mp}\right)^{\frac{-|m|-|s|}2} \left(z_{{}_\pm}\right)^{\frac{|m|-|s|}2}
\times \hskip 4truecm \\
\times\underline{\text{HeunC}}( 4\sigma_m a \omega,-|m|-|s|,|m|-|s|,- 4\sigma_m a\omega |s|,
{\frac{m^2+s^2}{2}}+ 2\sigma_m a\omega|s| -a^2\omega^2-E,z_{{}_\mp})
+\nonumber\\
+\,{}_s\Gamma^{\pm}_2(\omega,E,m)\, e^{ 2\sigma_m a \omega z_{{}_\mp}}
\left(z_{{}_\mp}\right)^{\frac{|m|+|s|}2} \left(z_{{}_\pm}\right)^{\frac{|m|-|s|}2}
\times \nonumber \hskip 3.8truecm \\
\times\text{HeunC}( 4\sigma_m a \omega,|m|+|s|,|m|-|s|,-4\sigma_m a\omega |s|,
{\frac{m^2+s^2}{2}}+ 2\sigma_m a\omega|s| -a^2\omega^2-E,z_{{}_\mp}).
\nonumber
\een

Now it is clear that in the general case, when ${}_s\Gamma^{\pm}_1(\omega,E,m)\!\neq\! 0$,
the solutions \eqref{TAE_delta_N_integer_S_A} are singular around the corresponding points $z_{{}_\pm}\!=\!1$
and in addition -- still infinite valued, because of the poles and of the
logarithmic terms in the concomitant confluent Heun function in Eq. \eqref{TAE_delta_N_integer_S_expansion},
as well as because of the singular factor $\left(z_{{}_\mp}\right)^{\frac{-|m|-|s|}2}$.
One can remove at once all these unwanted properties from solutions \eqref{TAE_delta_N_integer_S_expansion}
imposing the condition ${}_s\Gamma^{\pm}_1(\omega,E,m)\!=\!0$.
Unfortunately, the explicit form of the connection constants ${}_s\Gamma^{\pm}_{1,2}(\omega,E,m)$ is completely
unknown. At present, this is one of the main unsolved problems in the theory of the confluent Heun functions.

Another way to avoid the logarithmic terms in the solutions \eqref{TAE_delta_N_integer_S_A},
\eqref{TAE_delta_N_integer_S_expansion} is to impose the
$\Delta_{N+1}$-condition, reducing this way confluent Heun's functions to polynomials.
We consider in detail these two possibilities in the next sections 7 and 8.

2. The second class angular $\delta_N$-solutions: $\sigma_b=-\sigma_c$. Then we obtain
\ben
{}_sN_{m,\sigma_a,\sigma_b,-\sigma_b}+1= \pm\,m \sigma_b  -\sigma_a s \geq 1.
\la{sNm_Kerr_A}
\een

Now the additional requirement $\sigma_b(s\mp m)\geq 0$
and \eqref{sNm_Kerr_A} yield  the solutions
\ben
\la{TAE_delta_N_integer_S_A_II}
{}_sS_{\omega,E,m,-\sigma,\sigma,-\sigma}^{\pm}(z_{{}_\pm})=
\!e^{\mp 2\sigma a \omega z_{{}_\pm} }
\left(z_{{}_\pm}\right)^{\frac{|s|\mp\sigma m}2} \left(z_{{}_\mp}\right)^{\frac{-|s|\mp\sigma m}2}\times \hskip 6.4truecm\\
\times\text{HeunC}(\mp 4\sigma a \omega,|s|\mp\sigma m,-|s|\mp\sigma m, \pm 4\sigma a\omega|s|,{\frac{m^2+s^2}{2}} \mp 2\sigma a\omega|s| -a^2\omega^2-E,z_{{}_\pm}).
\nonumber
\een
with $|s|\geq 1/2$ and $m$ restricted
in the asymmetric finite intervals $1-|s|\leq \pm\sigma m\leq |s|$, i.e.,
$
$$ -|s|+(1\pm\sigma\sigma_m)/2 \leq |m|\leq|s|-(1\mp\sigma\sigma_m)/2.$$
$
These solutions have  $\sigma_a=-\sigma_b=\sigma_c=-\sigma$,
${}_sN_{m,-\sigma,\sigma,-\sigma}+1=|s| \pm\,\sigma m  \geq 1$
and correspond to pairs
$\{s,m\}$ in sectors I \eqref{sectorI:ab} and IV \eqref{sectorIII:ab}.
Under the above conditions the solutions ${}_sS_{\omega,E,m-\sigma,\sigma,-\sigma}^{\pm}(z_{{}_\pm})$
\eqref{TAE_delta_N_integer_S_A_II} are obviously regular around the points $z_{{}_\pm}=0$.
Their behavior around the second regular singular points $z_{{}_\pm}=1$
can be studied using the following expansion with respect to the corresponding local basis
\ben
\la{TAE_delta_N_integer_S_expansion_II}
{}_sS_{\omega,E,m,-\sigma,\sigma,-\sigma}^{\pm}(z_{{}_\pm})=
{}_s\Gamma^{\pm}_1(\omega,E,m,-\sigma,\sigma,-\sigma)\, e^{ \pm 2\sigma a \omega z_{{}_\mp}}
\left(z_{{}_\mp}\right)^{\frac{-|s|\mp\sigma m}2} \left(z_{{}_\pm}\right)^{\frac{|s|\mp\sigma m}2}
\times \hskip 1.9truecm \\
\times\underline{\text{HeunC}}( \pm4\sigma a \omega,-|s|\mp\sigma m,|s|\mp\sigma m,\mp 4\sigma a\omega |s|,
{\frac{m^2+s^2}{2}} \pm2\sigma a\omega|s| -a^2\omega^2-E,z_{{}_\mp})
+\nonumber\\
+\,{}_s\Gamma^{\pm}_2(\omega,E,m,-\sigma,\sigma,-\sigma)\, e^{ \pm 2\sigma a \omega z_{{}_\mp}}
\left(z_{{}_\mp}\right)^{\frac{|s|\pm\sigma m}2} \left(z_{{}_\pm}\right)^{\frac{|s|\mp\sigma m}2}
\times \nonumber \hskip .truecm \\
\times\text{HeunC}( \pm 4\sigma a \omega,|s|\pm \sigma m,|s|\mp \sigma m,\mp4\sigma a\omega |s|,
{\frac{m^2+s^2}{2}} \pm 2\sigma a\omega|s| -a^2\omega^2-E,z_{{}_\mp}).
\nonumber
\een
Since $-2|s|\leq-|s|\mp\sigma m\leq -1$ and $ 1\leq |s|\pm\sigma m\leq 2|s|$, the two independent local
solutions in \eqref{TAE_delta_N_integer_S_expansion_II} are well defined.
The first solution (with concomitant Heun's function) is singular around $z_{{}_\pm}=1$
($\Leftrightarrow z_{{}_\mp}=0$) and, in addition, infinite-valued.
One can remove at once all these unwanted properties from solutions \eqref{TAE_delta_N_integer_S_expansion_II}
imposing the condition  ${}_s\Gamma^{\pm}_1(\omega,E,m,-\sigma,\sigma,-\sigma)\!=\!0$.
Unfortunately, at present the explicit form of the connection constants
${}_s\Gamma^{\pm}_{1,2}(\omega,E,m,-\sigma,\sigma,-\sigma)$ is completely
unknown, too.

Another way to avoid the logarithmic terms in the solutions \eqref{TAE_delta_N_integer_S_A_II},
\eqref{TAE_delta_N_integer_S_expansion_II} is to impose the
$\Delta_{N+1}$-condition, reducing this way confluent Heun's functions to polynomials.
We consider in detail these two possibilities in the next sections 7 and 8.

As seen, in the case of the TAE the only role of the $\delta_N$-condition is to relate
the degree $N$ of the $\Delta_{N+1}$-condition with the spin-weight $s$ and the azimuthal number $m$
and to select the proper solutions.

Note that up to now only regular solutions to the TAE, which obey the condition \eqref{sN_Kerr_A}
have been studied and used in the literature \cite{Teukolsky,Chandra,angular}.
In section 7 we develop a new approach to the regular solutions,
based on confluent Heun's functions.
The nonregular angular $\delta_N$-solutions, subject to the  condition \eqref{sN_Kerr_A},
and the infinite series of solutions, subject to the condition \eqref{sNm_Kerr_A}, are introduced
and considered for the first time in the present paper.

\section{Regular solutions of the TAE}

The spectral conditions ${}_s\Gamma^{\pm}_1(\omega,E,m)= 0$ and ${}_s\Gamma^{\pm}_1(\omega,E,m,-\sigma,\sigma,-\sigma)=0$
ensure the regularity of the solutions \eqref{TAE_delta_N_integer_S_A} and \eqref{TAE_delta_N_integer_S_A_II},
as seen from the formulas  \eqref{TAE_delta_N_integer_S_expansion} and \eqref{TAE_delta_N_integer_S_expansion_II}.
One is not able to use these conditions directly, since the explicit form of the connection constants
${}_s\Gamma^{\pm}_1(\omega,E,m)$ and ${}_s\Gamma^{\pm}_1(\omega,E,m,-\sigma,\sigma,-\sigma)$ is not known.
Therefore, we are forced to use a roundabout way to find the regular solutions to the TAE.

Suppose we have a solution ${}_sS^{+\, reg}_{\omega,E,m}(\theta)$ which is regular around the S-pole ($\theta_S=\pi$) and
another solution  ${}_sS^{-\, reg}_{\omega,E,m}(\theta)$ which is regular around the N-pole ($\theta_N=0$).
We will have a solution ${}_sS^{\,\,{}_{REG}}_{\omega,E,m}(\theta)$, regular everywhere in the interval $\theta \in [0,\pi]$,
if and only if ${}_sS^{+\, reg}_{\omega,E,m}(\theta)\!=\!\text{const}\times {}_sS^{-\, reg}_{\omega,E,m}(\theta)$,
i.e., if the Wronskian vanishes:
$\text{W}\!\left[{}_sS^{+\, reg}_{\omega,E,m}(\theta),{}_sS^{-\, reg}_{\omega,E,m}(\theta)\right]\!=\!0$.
This condition determines  the constant $E$ in the form $E=E(a\omega,s,m,l)$, $l$ being a (half)integer.
The Wronskian will vanish for any $\theta\in [0,\pi]$, if it is zero for some $\theta_0\in (0,\pi)$.

To utilize this idea for all values of the parameters $s$ and $m$,
we have to divide the whole plane $\{s,m\}$ into four sectors and to choose
the solutions ${}_sS^{\pm\, reg}_{\omega,E,m}(\theta)$ defined by Eqs. \eqref{sectorI:ab}-\eqref{sectorIV:ab}.

The spectral condition makes equal the solutions of group (a) and the solutions of group (b) in each sector.
It can be written in different equivalent forms combining in pairs one solution from the group (a) and another one from the group (b).
Below we give the simplest form of this condition in each sector, written here for the first time in terms of confluent Heun's
function $\mathrm{HeunC}$ and its derivative $\mathrm{HeunC}^\prime$.
The set of all conditions \eqref{E_regular:abcd} defines the separation constant in the whole plane $\{s,m\}$ in the form
$E = (m^2+s^2)/2-a^2\omega^2+\varepsilon(a\omega, m, s)$. The
new parameter $\varepsilon(a\omega, m, s)$ is to be found from the following transcendental equations:
\begin{subequations}\label{E_regular:abcd}
\begin{align}
{{\mathrm{HeunC}^\prime(\pm 4 a\omega,\,s+m, s-m,-4 a\omega s,+2\omega as-\varepsilon,\left(\sin{{\theta}\over{2}}\right)^2)}
\over
{\mathrm{HeunC}(\pm 4 a\omega,\,s+m, s-m,-4 a\omega s,+2\omega as-\varepsilon,\left(\sin{{\theta}\over{2}}\right)^2)}}\,+
\hskip 3.3truecm\label{E_regular:a}\\
+\,{{\mathrm{HeunC}^\prime(\mp 4 a\omega,\,s-m, s+m,+4 a\omega s,-2a\omega s-\varepsilon,\left(\cos{{\theta}\over{2}}\right)^2)}
\over
{\mathrm{HeunC}(\mp 4 a\omega,\,s-m, s+m,+4 a\omega s,-2a\omega s-\varepsilon,\left(\cos{{\theta}\over{2}}\right)^2)}}
=0 \,\,\,\,\,\text{-- in sector I},\hskip .7truecm\nonumber
\end{align}
\begin{align}
{{\mathrm{HeunC}^\prime(\pm 4 a\omega,\,-s-m, s-m,-4 a\omega s,+2\omega as-\varepsilon,\left(\sin{{\theta}\over{2}}\right)^2)}
\over
{\mathrm{HeunC}(\pm 4 a\omega,\,-s-m, s-m,-4 a\omega s,+2\omega as-\varepsilon,\left(\sin{{\theta}\over{2}}\right)^2)}}\,+
\hskip 3.3truecm\label{E_regular:b}\\
+\,{{\mathrm{HeunC}^\prime(\mp 4 a\omega,\,s-m, -s-m,+4 a\omega s,-2a\omega s-\varepsilon,\left(\cos{{\theta}\over{2}}\right)^2)}
\over
{\mathrm{HeunC}(\mp 4 a\omega,\,s-m, -s-m,+4 a\omega s,-2a\omega s-\varepsilon,\left(\cos{{\theta}\over{2}}\right)^2)}}
=0 \,\,\,\,\,\text{-- in sector II},\hskip .5truecm\nonumber
\end{align}
\begin{align}
{{\mathrm{HeunC}^\prime(\pm 4 a\omega,\,-s-m, -s+m,-4 a\omega s,+2\omega as-\varepsilon,\left(\sin{{\theta}\over{2}}\right)^2)}
\over
{\mathrm{HeunC}(\pm 4 a\omega,\,-s-m, -s+m,-4 a\omega s,+2\omega as-\varepsilon,\left(\sin{{\theta}\over{2}}\right)^2)}}\,+
\hskip 3.3truecm\label{E_regular:c}\\
+\,{{\mathrm{HeunC}^\prime(\mp 4 a\omega,\,-s+m, -s-m,+4 a\omega s,-2a\omega s-\varepsilon,\left(\cos{{\theta}\over{2}}\right)^2)}
\over
{\mathrm{HeunC}(\mp 4 a\omega,\,-s+m, -s-m,+4 a\omega s,-2a\omega s-\varepsilon,\left(\cos{{\theta}\over{2}}\right)^2)}}
=0 \,\,\,\,\,\text{-- in sector III},\hskip .4truecm\nonumber
\end{align}
\begin{align}
{{\mathrm{HeunC}^\prime(\pm 4 a\omega,\,s+m, -s+m,-4 a\omega s,+2\omega as-\varepsilon,\left(\sin{{\theta}\over{2}}\right)^2)}
\over
{\mathrm{HeunC}(\pm 4 a\omega,\,s+m, -s+m,-4 a\omega s,+2\omega as-\varepsilon,\left(\sin{{\theta}\over{2}}\right)^2)}}\,+
\hskip 3.3truecm\label{E_regular:d}\\
+\,{{\mathrm{HeunC}^\prime(\mp 4 a\omega,\,-s+m, s+m,+4 a\omega s,-2a\omega s-\varepsilon,\left(\cos{{\theta}\over{2}}\right)^2)}
\over
{\mathrm{HeunC}(\mp 4 a\omega,\,-s+m, s+m,+4 a\omega s,-2a\omega s-\varepsilon,\left(\cos{{\theta}\over{2}}\right)^2)}}
=0 \,\,\,\,\,\text{-- in sector IV},\hskip .4truecm\nonumber
\end{align}
\end{subequations}
valid simultaneously for all  values of $\theta\in (0,\pi)$.
Thus, the two-singular-points boundary problem for the TAE is solved.
It yields a countable set of values $E(a\omega, m, s, l)$
numbered by some (half)integer $l$:
$l$ is  integer for an integer spin, or half-integer -- for a half-integer spin.
Due to the symmetries \eqref{symmetriesTAE16local:ab}
of the solutions to the TAE, the different relations \eqref{E_regular:a} and \eqref{E_regular:c}, or
\eqref{E_regular:b} and \eqref{E_regular:d} give similar results.
More precisely $E(a\omega, m, -s, l)=E(a\omega, m, s, l)$ and $E(-a\omega, -m, s, l)=E(a\omega, m, s, l)$.

An important consequence is that all the regular solutions obtained this way are angular $\delta_N$-solutions with
the same ${}_sN$ \eqref{sN_Kerr_A} in sectors II and IV, or with the same ${}_sN_{m,\sigma_a,\sigma_b}$ \eqref{sNm_Kerr_A}
-- in sectors I and III.
This is because between the solutions  \eqref{sectorII:ab} and \eqref{sectorIV:ab} we certainly have $\delta_N$-solutions:
${}_sS^{\pm}_{\omega,E,m,---}$, for $s>0$, and ${}_sS^{\pm}_{\omega,E,m,+++}$, for $s<0$.
Between the solutions  \eqref{sectorI:ab} and \eqref{sectorIII:ab} $\delta_N$-solutions are
${}_sS^{-}_{\omega,E,m,\mp\pm\mp}$, for $m>0$, and ${}_sS^{+}_{\omega,E,m,\mp\pm\mp}$, for $m<0$.
As a result of uniqueness of the regular solutions with given values of the parameters, all regular solutions inherit
the $\delta_N$-property.
Hence, all regular solutions of the TAE obey the Teukolsky-Starobinsly identities \cite{Fiziev2009b}.

Let us consider the limit $a\omega\to 0$ of the regular solutions to the TAE.
Since
\ben\text{HeunC}(0,\beta,\gamma,0,\eta,z)\!=\!
(1\!-\!z)^{\beta+\gamma+1+\sqrt{\beta^2+\gamma^2+1-4\eta}}\times \hskip 7.truecm\nonumber\\
{}_2F_{1}\left(
{\frac{\beta\!+\!\gamma\!+\!1+\!\sqrt{\beta^2\!+\!\gamma^2\!+\!1\!-\!4\eta}}{2}},
{\frac{\beta\!+\!\gamma\!+\!1-\!\sqrt{\beta^2\!+\!\gamma^2\!+\!1\!-\!4\eta}}{2}};
\beta\!+\!1;z\right)\!,
\la{HC_F}
\een
in this limit the Heun functions in Eqs. \eqref{sectorI:ab}-\eqref{sectorIV:ab} and \eqref{E_regular:abcd}
can be reduced to the Gauss hypergeometric ones.
Then, using the well-known properties of the Gauss hypergeometric function  ${}_2F_1$ one can derive
from Eqs. \eqref{E_regular:abcd} with $a\omega= 0$ the spectrum $E(0,s,l,m)=l(l+1)$,
$l=l(s,m,\bar l)=\max(|m|,|s|)+\bar l,\,\, \bar l=0,1,2,\dots$
The values of the separation constant $E(0,s,l,m)$ in this case are real.
The numerical analysis of Eqs. \eqref{E_regular:abcd}
written directly in terms of confluent Heun's functions
confirms this standard result for the  limit $a\omega\!=\!0$.
The corresponding regular confluent Heun functions in ${}_sS^{\,{}_{REG}}_{\omega=0,l,m}(\theta)$
in this case are reduced to Jacobi's polynomials -- in the case of an integer spin,
and to their spin-weighted generalizations -- for a half-integer spin \cite{HalfInteger}.

The solutions $E(a\omega,s,l,m)$ for small $a\omega$ and integer spin have been studied
many times \cite{Teukolsky,angular} in the form of Taylor's series
expansion $E(a\omega,s,l,m)\!=\!l(l\!+\!1)\!+\!\sum_{j=\!1}^\infty
E_{j,s,l,m}(a\omega)^j$ without use of Eqs. \eqref{E_regular:abcd}
and without utilizing the Heun functions.
A little bit surprising thing is that the solutions ${}_sS^{\,{}_{REG}}_{\omega,l,m}(\theta)$ with $a\omega\neq 0$,
regular at both poles, are not polynomial and can be represented as an infinite series with respect to Jacobi's polynomials.
Here we describe the regular solutions to the TAE in terms of confluent $\delta_N$-Heun's functions for the first time.

\section{Polynomial Solutions of the TAE}

\subsection{Singularities of the Polynomial solutions to the TAE}
The polynomial solutions to the TAE are a special subclass of the angular $\delta_N$-solutions studied in section 6,
since both of the two conditions \eqref{PolynomCond:ab} are valid for them.
Being a polynomial in $z$, the HeunC-factor
is regular at both regular singular points $\theta=0,\pi$.
Then the singularities of the polynomial solutions
around the poles are defined completely by the factors
$\left(z_{{}_\pm}\right)^{\sigma_b{{b_{{}_\pm}}/ 2}}$ and $\left(z_{{}_\mp}\right)^{\sigma_c{{c_{{}_\pm}}/ 2}}$
in Eq. \eqref{TAE16local}. Thus:

1. In case of the first class  angular $\delta_N$-solutions \eqref{TAE_delta_N_integer_S_A} with
$$|m|\geq|s|$$
we see that the singularities are defined by the factor $\left(z_{{}_\mp}\right)^{\sigma_c{{c_{{}_\pm}}/ 2}}$
which gives ${}_sS_{\omega,E,m}^{+}(z_{{}_\pm})\sim \left(\sin{\theta\over 2}\right)^{-(|s|+|m|)}$, i.e., singularity at the N-pole $\theta=0$,
and ${}_sS_{\omega,E,m}^{-}(z_{{}_\pm})\sim \left(\cos{\theta\over 2}\right)^{-(|s|+|m|)}$, i.e., singularity at the S-pole $\theta=\pi$.

2. In case of the second class angular $\delta_N$-solutions \eqref{TAE_delta_N_integer_S_A_II} with
$$ -|s|+(1\pm\sigma\sigma_m)/2 \leq |m|\leq|s|-(1\mp\sigma\sigma_m)/2$$
we see that the singularities are defined by the factor $\left(z_{{}_\mp}\right)^{\sigma_c{{c_{{}_\pm}}/ 2}}$
which gives ${}_sS_{\omega,E,m,-\sigma,\sigma,-\sigma}^{+}(z_{{}_\pm})\sim \left(\sin{\theta\over 2}\right)^{-(|s|+\sigma m)}$,
i.e., singularity at the N-pole $\theta=0$, and
${}_sS_{\omega,E,m,-\sigma,\sigma,-\sigma}^{-}(z_{{}_\pm})\sim \left(\cos{\theta\over 2}\right)^{-(|s|-\sigma m)}$,
i.e., singularity at the S-pole $\theta=\pi$.

As a result, we see that in any case the polynomial solutions are regular around one of the poles
and singular around the other one.

Using relations (\ref{munu:b}) and  (\ref{TAEparameters}) we obtain the general formula for the constant $E$ in the form
\ben
E^\pm\!=\!\mu^\pm\!-\!a\omega^2 \mp 2\sigma_a \big(1\!\mp\sigma_b m+(\sigma_a+\sigma_b)s\big) a\omega +
{\frac{\sigma_b\!-\sigma_c}{2}}m\left(\sigma_b m \mp 1\right) +
{\frac{\sigma_b\!+\sigma_c}{2}}s\left(\sigma_b s\!+1\right).
\la{E_TAE}
\een

Further analysis shows that some of the properties of the two classes of polynomial solutions to the TAE
resemble the corresponding properties of the two classes of polynomial solutions of the TRA,
but there exist also some essential differences.

\subsection{First Class of Polynomial Solutions to the TAE:}

These are the solutions ${}_sS_{\omega,E,m,-\sigma,-\sigma,-\sigma}^{\pm}$ with $\Delta_{N+1}$-condition fulfilled.
For them the specific requirement $\sigma_b b_{{}_\pm}\geq 0$ yields the restriction $|m|\geq|s|$ and
the condition (\ref{sN_Kerr_A}) is fulfilled independently of the values of the azimuthal number $m$.
As in the case of the first class polynomial solutions to the TRA -- section 4, the value $s=0$ is eliminated
by (\ref{sN_Kerr_A}).
Hence, we have an infinite series of the first class polynomial solutions to the TAE for all admissible values of $s$ and $m$.
Preserving the style accepted in the previous sections we denote the polynomial
solutions to the TAE of the first class as ${}_sS_{\omega,E,m}^{\pm}={}_sS_{\omega,E,m,-\sigma,-\sigma,-\sigma}^{\pm}$.

For them the $\Delta_{N+1}$-condition reads $\Delta_{2|s|}(\mu)=0$ and has $2|s|$-in-number
solutions ${}_s\mu^\pm_{\omega,k,m}$. From formulae (\ref{E_TAE}) one obtains
\ben
{}_s E^\pm_{\omega,k,m}={}_s\mu^\pm_{\omega,k,m} +|s|(|s|-1) -a\omega(a\omega-2m) \mp 2\sigma(2|s|-1)a\omega,
\la{E_first_classA}
\een
where $k=1,\dots,2|s|$, $s=\pm 1/2,\pm 1, \pm 3/2, \pm 2$ and in addition $|m|\geq|s|$.

Solving the $\Delta_{N+1}$-condition, we obtain for the different values of $|s|$:
\ben
{}_sE_{\omega,m}^{\pm}=-a^2\omega^2+2a\omega m-{\frac 1 4}:\,\,\,\text{for}\,\,|s|={\frac 1 2},\, m=\pm 1/2, \pm 3/2,\dots;
\la{E_first_class_1/2_A}
\een
\ben
{}_sE_{\omega,k,m}^{\pm}=-a^2\omega^2+2a\omega \left(m-(-1)^k \sqrt{1-m/a\omega}\right):\,\,\,\text{for}\,\,k=1,2;\,|s|=1,\,\,m=\pm 1, \pm 2, \dots
\la{E_first_class_1_A}
\een

The values \eqref{E_first_class_1/2_A} and \eqref{E_first_class_1_A} of the separation constant $E$
obtained for the first class polynomial solutions to the TAE are the same as
the corresponding values \eqref{E_first_class_1/2} and \eqref{E_first_class_1}
for the first class polynomial solutions to the TRE.
Important consequences of this unexpected fact are considered in a separate paper \cite{RBPF}.

For the gravitational waves ($|s|=2$) the quantities
${}_s\mu_{\omega,k,m}^{\pm}$ are solutions of the algebraic equations of the fourth degree $\Delta^\pm_4(\mu)=0$.
We do not need here the exact form of these roots. It is quite complicated.
Below we present only the form of the separation constant $E$ for the TAE obtained
making use of the Taylor series expansions of the roots around the point $a\omega=0$.

Thus, we obtain for $|s|=2$, $k=1,2$, and $m=\pm 2, \pm 3, \dots$ the following eight series of values:
\ben
{}_sE_{\omega,k,m}^{\pm}=2-4ma\omega -i(-1)^k12\sqrt{(m-1)m(m+1)}\,(a\omega)^{3/2}
+6\left(m^2-{\frac 7 6}\right)(a\omega)^2
\!+\!{\cal{O}}_{5/2}(a\omega),\hskip .truecm\label{mu:1_2A}
\een
and for $|s|=2$, $k=3,4$, and $m=\pm 2, \pm 3, \dots$ another eight series of values:
\ben
{}_sE_{\omega,k,m}^{\pm}\!=\!-(-1)^k4\sqrt{m a\omega}\left(1+\left(3m-{\frac 2 m}\right)a\omega+{\cal{O}}_{2}(a\omega)\right)\!
+\!8ma\omega\!-\!6\!\left(\!m^2\!-\!{\frac 5 6}\!\right)\!(a\omega)^2\!+\!{\cal{O}}_{3}(a\omega).
\hskip .5truecm\label{mu:3_4A}
\een

As seen, for gravitational waves of the first polynomial class the values \eqref{mu:1_2A} and \eqref{mu:3_4A}
of the corresponding constants $E$ differ substantially from the analogous values \eqref{mu:1_2} and \eqref{mu:3_4}
of the constants $E$ obtained for the TRE in section 4.1.1.
This is in sharp contrast to the case of neutrino waves ($|s|=1/2$) of the first polynomial class
and to the case of electromagnetic waves ($|s|=1$) of this kind.

It can be shown that this phenomenon reflects the difference between the Starobinsky constants
for solutions with spin $2$ to the TAE and for solutions with the same spin $2$ to the TRE \cite{Teukolsky, Chandra, Fiziev2009b}.
The solutions to the TAE and to the TRE with the same spin $1/2$ or $1$ have the same Starobinsky constants.

Despite the above essential difference, the first-polynomial-class-solutions to the TAE and to the TRE
with spin $2$ have similar qualitative properties, discussed at the end of section 4.1.1.

\subsection{Second Class of Polynomial Solutions to the TAE:}
We have a finite number of second class polynomial solutions to the TAE for which the relation $\sigma_c=-\sigma_b$ holds.
For brevity, we list here only the ones of spin 2, 1 and 1/2.
For them the conditions $N\!\geq\!0$ and $\sigma_b b_{{}_\pm}\!\geq\!0$ must be satisfied simultaneously,
yielding the requirement $-|s|\leq-|s|+(1\pm\sigma\sigma_m)/2\leq|m|\leq|s|-(1\mp\sigma\sigma_m)/2\leq |s|$
-- almost opposite to the analogous requirement $|m|\geq |s|$ for the polynomial solutions of the first class.
Altogether there exist only the following 32 polynomial solutions of the second class ${}_sS_{\omega,E,m,\mp,\pm,\mp}^{\pm}$
with spin 2, 1 and 1/2:
\begin{eqnarray}
&{}_sS_{\omega,E,m,-,+,-}^{+}:&   s\!=\!+2,\,\, m\!=\!-1,\,\,\,\,0,\,1,\,2;\,\,  s\!=\!+1,\,\, m\!=\!\,\,\,\,0,\,1;\,\, s\!=+1/2,\,\,m\!=+1/2,-1,2;\nonumber \\
&{}_sS_{\omega,E,m,+,-,+}^{+}:&   s\!=\!-2,\,\, m\!=\!-2,-1,\,0,\,1;\,\,         s\!=\!-1,\,\, m\!=\!-1, 0;  \,\, s\!=-1/2,\,\,m\!=-1/2,+1/2;\nonumber\\
&{}_sS_{\omega,E,m,-,+,-}^{-}:&   s\!=\!+2,\,\, m\!=\!-2, -1,\,0,\,1;\,\,        s\!=\!+1,\,\, m\!=\!-1,\,0; \,\, s\!=+1/2,\,\,m\!=+1/2,-1/2;\nonumber\\
&{}_sS_{\omega,E,m,+,-,+}^{-}:&   s\!=\!-2,\,\, m\!=\!-1,\,\,\,\,0,\,1,\,2;\,\,  s\!=\!-1,\,\, m\!=\!\,\,\,\,\,0,\,1,\,\, s\!=-1/2,\,\,m\!=-1/2,+1/2.
\label{Second_Class_Polynomial_TAE}
\end{eqnarray}
The relation between the constants $E$ and $\omega$ follows from \eqref{E_TAE},
when $\mu$ in it is replaced by the solutions
of the $\Delta_{N+1}$-condition in the form $\Delta_{|s\pm m|}^\pm(\mu)=0$.
Here we omit these relations.

\section{The 256 Classes of Exact Factorized Solutions \\ to the Teukolsky Master Equation}
Combining solutions to the TRE and to the TAE studied in the previous sections
we can construct the following 256 classes of exact factorized solutions to the TME
\ben
{}_s{\cal K}^{\pm,\pm}_{\omega,E,m,\sigma_\alpha,\sigma_\beta,\sigma_\gamma,\sigma_a,\sigma_b,\sigma_c}(t,r,\theta,\varphi)=
e^{-i\omega t} e^{ im\varphi}
{}_sR^\pm_{\omega,E,m,\sigma_\alpha,\sigma_\beta,\sigma_\gamma}(r;r_{+},r_{-}) {}_sS_{\omega,E,m,\sigma_a,\sigma_b,\sigma_c}^{\pm}(\theta).
\la{256_Sol_TME}
\een

For specific physical problems one has to impose specific additional conditions,
like stability conditions, boundary conditions,
casuality conditions, specific fixing of the in-out properties, regularity conditions etc.
Thus, one selects some specific combinations of solutions
to the TRE and to the TAE  in Eq. \eqref{256_Sol_TME} and derives
the spectrum of the separation constants $\omega$ and $E$ in the given problem.

For example, choosing solutions to the TRE which enter both the event horizon and the 3D-space infinity we
study the Kerr black holes (for $a<M$), or necked singularities (for $a>M$) \cite{Teukolsky, Chandra}.
If in addition we choose regular solution to the TAE, we will obtain
the standard QNM of the Kerr black holes, or necked singularities.
Using in Eq. \eqref{256_Sol_TME} other solutions to the TRE and/or to the TAE, we may hopefully
describe different physical objects and phenomena,
for example, collimated jets, see in \cite{PFDS,PFDS_astroph:HE}.

The solutions \eqref{256_Sol_TME} do not necessarily have a direct physical meaning,
see the Introduction.
Instead, some linear combination of the specific solutions, which obey proper boundary conditions,
is to describe the Nature. In general the solutions \eqref{256_Sol_TME} have to be considered
as auxiliary mathematical objects -- (maybe singular) kernels of integral representations
\eqref{IntRepresentation_E_omega} of the physical solutions.
The choice of the corresponding amplitudes ${}_sA_{\omega,E,m,\sigma_\alpha,\sigma_\beta,\sigma_\gamma,\sigma_a,\sigma_b,\sigma_c}$
will fix completely the physical model and may ensure the convergence of the integrals and discrete sums
to physically acceptable solutions.
Since at present we have no rigorous mathematical treatment of this complicated issue,
we will study it in the next section using some constructive examples.

\section{Construction of Bounded Linear Combinations of Polynomial Solutions to the TAE}

We have seen in section 8 that the polynomial solutions to the TAE
are singular and unbounded with respect to the angle $\theta$ around the N-pole, or around  S-pole.
These solutions produce
a singular kernel in the integral representation \eqref{IntRepresentation_E_omega}.
It is important to know whether it is possible to have bounded with respect to the angle $\theta\in[0,\pi]$
solutions  ${}_s\Psi(t,r,\theta,\varphi)$ defined by Eq. \eqref{IntRepresentation_E_omega},
despite the singular character of the kernel in it.
The answer to this question is a quite nontrivial issue.
Here we reach a positive answer for perturbations of spin $1/2$ in several steps.

Let us consider the simplest case of double polynomial solutions of the first class to the TME with spin $1/2$ and $s=\sigma/2$.
For them we have an essential simplification, since according to Eqs. \eqref{sN_Kerr} and \eqref{sN_Kerr_A} ${}_sN=0$.
Hence, the HeunC-factors in both the radial and the angular polynomial solutions
are equal to $\text{const}\equiv 1$.
The value of the separation constant $E=-a^2\omega^2+2a\omega m-{\frac 1 4}$ is uniquely defined in both cases
by Eqs. \eqref{E_first_class_1/2}  and \eqref{E_first_class_1/2_A}.
Hence, the integration over the constant $E$ in \eqref{IntRepresentation_E_omega}
produces only one term with this fixed value.
As a result, the corresponding singular kernel \eqref{256_Sol_TME} is\footnote{To simplify
formula \eqref{K1:2}, we have omitted some constant factors in the corresponding solutions to the TRE and  TAE,
which do not depend continuously on the {\em real} variables $r$ and $\theta$, but may have different values outside the event horizon,
in the domain between the event horizon and the Cauchy horizon and inside the Cauchy horizon. This is a legal operation, since one can
include these factors in the amplitudes ${}_sA_{\dots}$ in the representation \eqref{IntRepresentation_E_omega},
considering separately the different domains where these factors are constant.}:
\ben
{}_{\frac \sigma 2}{\cal K}_{\omega,E,m}(t,r,\theta,\varphi)=\delta\left(E+a^2\omega^2-2ma\omega+1/4\right)\Delta^{-\frac{1+\sigma}4} e^{-i\omega T_\sigma}{\frac{\left(W_\sigma\right)^m}{\sqrt{\sin\theta}}},
\la{K1:2}
\een
where $T_\sigma=t+\sigma\left(r_*-ia\cos\theta\right)$, $W_\sigma=e^{i\phi_\sigma}\cot{{\theta_\sigma}\over 2}$, $\phi_\sigma=\varphi+{\sigma\over {2p}}\ln\left|{\frac{r-r_{+}}{r-r_{-}}}\right|$, and $\theta_\sigma\!=\!\theta$, if $\sigma\!=\!+1$, or
$\theta_\sigma\!=\!\pi-\theta$, if $\sigma=-1$.

The complex variable $W_\sigma$ defines a stereographic projection of the two-sphere $\mathbb{S}^{(2)}_{\phi_\sigma,\theta_\sigma}$ on the compactified complex plane $\mathbb{\tilde C}_{W_\sigma}$.
Its use is critical for further analysis of the problem. Note that after the transition from real variables $\{\theta, \phi_\sigma\}$
to the complex one $W_\sigma$ one must introduce an additional phase factor $\exp(-is\phi_\sigma)$ in the spin-weighted spheroidal harmonics,
due to the back rotation of the basis (See the paper by Goldberg et al. in \cite{HalfInteger}.).
In the case of spin $1/2$ the introduction of such factor $\exp(\mp i\phi_\sigma/2)$ is equivalent to a transition
in what follows from half-integer to integer values of the azimuthal number $m$ and
a replacement $m\rightarrow m\pm1/2$ in the factor $\delta\left(E+a^2\omega^2-2ma\omega+1/4\right)$ in Eq. \eqref{K1:2}.

Taking the trivial integral on the variable $E$, one obtains from the representation \eqref{IntRepresentation_E_omega} and Eq. \eqref{K1:2}
\begin{equation}\la{IntRepresentation_polynom__omega_s1:2}
{}_{\sigma\over 2}\Psi(t,r,\theta,\varphi)=\Delta(r)^{-\frac{1+\sigma}4}
\sqrt{\left({|W_\sigma|+|W_\sigma|^{-1}}\right)/ 2}\sum_{m=-\infty}^{\infty}
\left({1\over 2\pi} \int\limits_{\mathcal{L_\omega}}\!\!d\omega\,\,e^{-i\omega T_\sigma}\,{}_{\sigma\over 2}A_{\omega,m}\right)
\left(W_\sigma\right)^m.
\end{equation}
Since in this case we have no other restriction on the frequencies $\omega$, different from the stability requirement $\Im(\omega)<0$,
the otherwise arbitrary integration contour ${\mathcal{L_\omega}}\in \mathbb{C}_\omega$
in \eqref{IntRepresentation_polynom__omega_s1:2} must lie in the
lower complex half-plane. Suppose that the amplitudes ${}_{\sigma\over 2}A_{\omega,m}$ and the contour ${\mathcal{L_\omega}}$
are chosen in such way that for all $m\in \mathbb{Z}$ there exist well defined integrals
\ben
\la{AT}
{1\over 2\pi} \int\limits_{\mathcal{L_\omega}}\!\!d\omega\,\,e^{-i\omega T_\sigma}\,{}_{\sigma\over 2}A_{\omega,m}=
{}_{\sigma\over 2} \mathfrak{A}_m(T_\sigma).
\een
Then
\begin{equation}\la{Sum_polynom__omega_s1:2}
{}_{\sigma\over 2}\Psi(t,r,\theta,\varphi)=\Delta(r)^{-\frac{1+\sigma}4}
\sqrt{\left({|W_\sigma|+|W_\sigma|^{-1}}\right)/ 2}\sum_{m=-\infty}^{\infty}
{}_{\sigma\over 2}\mathfrak{A}_m\left(T_\sigma\right)
\left(W_\sigma\right)^m.
\end{equation}

Suppose, in addition, that in some ring domain $|W_\sigma|\!\in\!\left(|W|^\prime, |W|^{\prime\prime}\right)$,
$0\!<\!|W|^\prime\!<\!|W|^{\prime\prime}\!<\!\infty$ the sum $\sum_{m=-\infty}^{\infty}
{}_{\sigma\over 2}\mathfrak{A}_m\left(T_\sigma\right)\left(W_\sigma\right)^m=
{}_{\sigma\over 2}\mathfrak{A}\left(T_\sigma,W_\sigma\right)$
represents a convergent Laurent series of some analytic function
${}_{\sigma\over 2}\mathfrak{A}\left(T_\sigma,W_\sigma\right)$.
For this purpose the coefficients ${}_{\sigma\over 2}\mathfrak{A}_m\left(T_\sigma\right)$ in Eq. \eqref{AT} for $m>0$
and, independently, for $m<0$ must satisfy some of the well-known criteria for convergence of the
corresponding series.
Thus, we finally obtain a solution to the TME with spin $1/2$
which depends on an arbitrary analytic
function  ${}_{\sigma\over 2}\mathfrak{A}\left(T_\sigma,W_\sigma\right)$
of the two variables $T_\sigma$ and $W_\sigma$:
\begin{equation}\la{General_1:2}
{}_{\sigma\over 2}\Psi(t,r,\theta,\varphi)=\Delta(r)^{-\frac{1+\sigma}4}
\sqrt{\left({|W_\sigma|+|W_\sigma|^{-1}}\right)/ 2}\,\,
{}_{\sigma\over 2}\mathfrak{A}\left(T_\sigma,W_\sigma\right).
\end{equation}
Returning to the Boyer-Lindquist variables one can check directly that \eqref{General_1:2}
indeed gives a general solution to the TME with spin $1/2$. The explicit form of the variable $T_\sigma$ shows that
outside the event horizon these solutions describe {\em one-way-running} waves:
outgoing to space infinity running waves -- for $\sigma=-1$ and incoming from space infinity running waves -- for $\sigma =+1$.

Now it is easy to remove the singularities from the $z$-axis, i.e. on the poles $\theta =0,\pi$. For example, let us choose
${}_{\sigma\over 2}\mathfrak{A}\left(T_\sigma,W_\sigma\right)=1/\sqrt{\left(W_\sigma+W_\sigma^{-1}\right)/2}$.
Then ${}_{\sigma\over 2}\Psi(t,r,\theta,\varphi)=\Delta(r)^{-\frac{1+\sigma}4}/{\sqrt{1-\sin^2\phi_\sigma\sin^2\theta_\sigma}}$
has no singularities on the poles  $\theta_\sigma =0,\pi$,
but this way we have worked out two new singular lines
$\phi_\sigma=\varphi+{\sigma\over {2p}}\ln\left|{\frac{r-r_{+}}{r-r_{-}}}\right|=\pm\pi/2$
on the equatorial plane $\theta=\pi/2$.
Hence, this way the singular line of the solution has been only deformed and translated to a new position.
The same happens if we choose the more general function
${}_{\sigma\over 2}\mathfrak{A}\left(T_\sigma,W_\sigma\right)=
1/\sqrt{\left( a(T_\sigma) W_\sigma + b(T_\sigma) W_\sigma^{-1} + c(T_\sigma) \right)/2}$.
In this case, the singular $z$-axis will be deformed, translated and doubled to the non-static singular lines
$\phi_\sigma=\varphi+{\sigma\over {2p}}\ln\left|{\frac{r-r_{+}}{r-r_{-}}}\right|=\phi_{1,2}=\arg(W_{1,2})$
on the (in general) moving cones $\theta\!=\!\theta_{1,2}\!=\!\arctan\left(|W_{1,2}|^{-1}\right)$,
where $W_{1,2}$ are the two roots of the equation $a(T_\sigma) W_\sigma\!+\!b(T_\sigma) W_\sigma^{-1}\!+\!c(T_\sigma)\!=\!0$.
Here we have chosen a special form of the function ${}_{\sigma\over 2}\mathfrak{A}\left(T_\sigma,W_\sigma\right)$
which yields finite nonzero values of the solution on the poles $\theta =0,\pi$.

It is possible to chose the function ${}_{\sigma\over 2}\mathfrak{A}\left(T_\sigma,W_\sigma\right)$
with the denominator which is a sum of polynomials of higher degree with respect to variables $W_\sigma$ and $W_\sigma^{-1}$.
Then the solution \eqref{General_1:2} equals zero at the N and S-poles and
we can work out an arbitrary number of singular lines of the solution
related to the zeros of the denominator.
At first glance, this possibility may not seem to be interesting for the physical applications,
since on the singular lines the linear perturbation theory in use is not applicable.
We mention it here just to have a clear mathematical picture.
It is interesting to study the same situation in the whole nonlinear theory and to know whether in it
the singular lines may be replaced by regular ones. If so, the perturbation theory under consideration
indicates a possible complicated structure of the exact radiation field on the Kerr background.

The most important question for a correct application of the linear perturbation theory under consideration,
is whether one can find a regular analytical function
${}_{\sigma\over 2}\mathfrak{A}\left(T_\sigma,W_\sigma\right)$
without singularities in the complex plane $\mathbb{C}_{W_\sigma}/\{0,\infty\}$, i.e.,
with the points $W_\sigma=0$ and $W_\sigma=\infty$ punctured and which, in addition,
can remove the unbounded increase of the solutions due to the singularities of the
factor $\sqrt{\left({|W_\sigma|+|W_\sigma|^{-1}}\right)/ 2}$ in \eqref{General_1:2}.
We give a positive answer to this question constructing two explicit examples:

1. Using the basic equality
${\sum\limits_{m=-\infty}^{\infty}} W^m I_m(z)\!=\!\exp\left({1\over 2}\left(W\!+\!W^{-1}\right)z\right)$
for the modified Bessel functions $I_m(z)$ \cite{BE} we choose the coefficients in \eqref{Sum_polynom__omega_s1:2}
in the specific form ${}_{\sigma\over 2}\mathfrak{A}_m\left(T_\sigma\right)=
\exp\left(-{\bar\sigma\over 2}\omega^2T_\sigma^{\,2}\right)I_m(\omega T_\sigma)$, where
$\omega=\omega_R+i\omega_I$ is a fixed frequency and $\bar\sigma=\text{sign}(|\omega_R|-|\omega_I|)$.
Then
\begin{equation}\la{Regular_1:2_1}
{}_{\sigma\over 2}\Psi_\omega(t,r,\theta,\varphi)=\Delta(r)^{-\frac{1+\sigma}4}
\sqrt{\left({|W_\sigma|+|W_\sigma|^{-1}}\right)/ 2}\,\,
\exp\left(-{\bar\sigma\over 2}\omega^2T_\sigma^{\,2}\right)\exp\left({1\over 2}\left(W_\sigma\!+\!W_\sigma^{-1}\right)\omega T_\sigma\right)
\end{equation}
is a stable solution, since by construction it goes to zero when $t\to +\infty$.
It is not difficult to obtain its limit when $\theta_\sigma \to 0,\pi$
in the form
\ben
\la{Regular_1:2_Limit_1}
\lim\limits_{\theta_\sigma \to 0,\pi}\left({}_{\sigma\over 2}\Psi_\omega(t,r,\theta,\varphi)\right)=
\Delta(r)^{-\frac{1+\sigma}4}
\exp\left(-{\bar\sigma\over 2}\omega^2 T_{\sigma;0,\pi}^2\right)\times
\hskip 3.4truecm\nonumber\\
\times\lim\limits_{\theta_\sigma \to 0,\pi}\left({1\over{\sqrt{\sin\theta}}}
\exp\left(
{\frac
{|\omega|\sqrt{(t+\sigma r_*)^2+a^2}}
{\sin\theta}e^{i\Upsilon_{\omega\sigma;0,\pi}}}
\right)\right).
\een
Here
\ben
\la{Upsilon}
\Upsilon_{\omega,\sigma;0,\pi}=
\pm\Bigg(\varphi+{\sigma\over {2p}}\ln\left|{\frac{r-r_{+}}{r-r_{-}}}\right|-
\sigma\arctan\left({\frac{a}{t+\sigma r_*}}\right)\Bigg)+\arg (\omega),
\,\,\,\text{for}\,\,\,\theta=0,\,\,\text{or}\,\,\pi
 \een
is the limit of the total phase of the term ${1\over 2}\left(W_\sigma\!+\!W_\sigma^{-1}\right)\omega T_\sigma$ and
$T_{\sigma;0,\pi}=t+\sigma(r_*\mp ia)$.
In Eq. \eqref{Upsilon} the sign $(+)$ corresponds to the limit $\theta_\sigma\to 0$ and the sign $(-)$ --
to the limit $\theta_\sigma\to \pi$.
Formula \eqref{Regular_1:2_Limit_1} shows that when $\Upsilon_{\omega,\sigma;0,\pi}\in\left(-{\pi\over 2},{\pi\over 2}\right)$
the solution ${}_{\sigma\over 2}\Psi_\omega(t,r,\theta,\varphi)$ is bounded everywhere in the interval $\theta\in [0,\pi]$,
since in this case $\lim\limits_{\theta_\sigma \to 0,\pi}\left({}_{\sigma\over 2}\Psi_\omega(t,r,\theta,\varphi)\right)=0$.
Otherwise this limit diverges and the solution is singular and unbounded around the poles.

Actually, the value of the parameter $\Upsilon_{\omega,\sigma;0,\pi}$ is not defined from a geometrical point of view,
because the value of the angle $\varphi$ is completely arbitrary on the poles $\theta_\sigma =0,\pi$.
As a result, we can choose any value of the parameter $\Upsilon_{\omega,\sigma;0,\pi}$
without changing the geometrical points
associated with the N and S-poles of the sphere $\mathbb{S}^{(2)}_{\theta,\varphi}$.
Since the different values of this parameter yield different solutions of the TAE,
we see that under the boundary conditions at hand the corresponding differential operator
is not self-adjoint \cite{RS}, but its self-adjoint extensions
do exist and can be fixed by suitable fixing of the free parameter $\Upsilon_{\omega,\sigma;0,\pi}$.
An analogous phenomenon is well known for the potentials $V(x)\sim$ $1/x^2$, or $1/r^2$ in quantum mechanics \cite{RS}.
Note that around the poles $\theta_\sigma =0,\pi$
the potential in the TAE \eqref{angE:ab} has precisely the same behavior:
${}_sW_{\omega,E,m}(\theta) \sim 1/\theta^2$ for $\theta\to 0$, and
${}_sW_{\omega,E,m}(\theta) \sim 1/(\theta-\pi)^2$ for $\theta\to \pi$.
In our case, the fixing of the parameter $\Upsilon_{\omega,\sigma;0,\pi}\in\left(-{\pi\over 2},{\pi\over 2}\right)$
makes the solutions \eqref{Regular_1:2_Limit_1} to the TME for spin $1/2$ smooth and bounded everywhere in the interval
$\theta\in [0,\pi]$, i.e., physically acceptable.

2. Another solution, which is {\em finite} everywhere in the interval $\theta\in [0,\pi]$
but has an infinite number of bounded oscillations around the poles $\theta=0,\pi$
can be obtained using the following equality
for the Bessel functions $J_m(z)$:
${\sum\limits_{m=-\infty}^{\infty}} (-1)^m W^{2m} \left(J_m(z)\right)^2\!=\!J_0\Big(\left(W\!+\!W^{-1}\right)z\Big)$
\cite{BE}. Now we choose the coefficients in \eqref{Sum_polynom__omega_s1:2}
in the specific form ${}_{\sigma\over 2}\mathfrak{A}_{2m}\left(T_\sigma\right)=
(-1)^m\exp\left(-{\bar\sigma\over 2}\omega^2T_\sigma^{\,2}\right)\left(J_{2m}(\omega T_\sigma)\right)^2$ and
${}_{\sigma\over 2}\mathfrak{A}_{2m+1}\left(T_\sigma\right)=0$ using the same notation as in the previous example. Then
\begin{equation}\la{Regular_1:2_2}
{}_{\sigma\over 2}\Psi_\omega(t,r,\theta,\varphi)=\Delta(r)^{-\frac{1+\sigma}4}
\sqrt{\left({|W_\sigma|+|W_\sigma|^{-1}}\right)/ 2}\,\,
\exp\left(-{\bar\sigma\over 2}\omega^2T_\sigma^{\,2}\right)J_0\Big(\left(W_\sigma\!+\!W_\sigma^{-1}\right)\omega T_\sigma\Big)
\end{equation}
is a stable solution to the TME with spin $1/2$. Taking into account the asymptotic expansion of the Bessel
function $J_0(z)\sim \sqrt{{\frac{2}{\pi z}}}\cos(z-\pi/4)$ we obtain in the limits $\theta_\sigma\to 0,\pi$:
\ben
\la{Regular_1:2_Limit_2}
\lim\limits_{\theta_\sigma \to 0,\pi}\left({}_{\sigma\over 2}\Psi_\omega(t,r,\theta,\varphi)\right)=
\Delta(r)^{-\frac{1+\sigma}4}
\exp\left(-{\bar\sigma\over 2}\omega^2 T_{\sigma;0,\pi}^{\,2}\right)
{1\over{\sqrt{\pi\omega T_{\sigma;0,\pi}}}}\times
\hskip 2.5truecm\nonumber \\
\times\lim\limits_{\theta_\sigma \to 0,\pi}
\left(\cos\left(
{\frac
{2|\omega|\sqrt{(t+\sigma r_*)^2+a^2}}
{\sin\theta}e^{i\Upsilon_{\omega\sigma;0,\pi}}}
\right)\right).
\een
As seen from Eq. \eqref{Regular_1:2_Limit_2}, there exist only two choices of the free parameter:
$\Upsilon_{\omega,\sigma;0,\pi}=0,\pi$, for which the solutions \eqref{Regular_1:2_2}
are finite everywhere in the interval $\theta\in [0,\pi]$ -- a critical property
for the use of the linear perturbation theory.
Approaching these poles the solutions oscillate infinitely many times with bounded finite amplitudes.
In this sense, the N and S-poles remain singularities of the bounded solutions \eqref{Regular_1:2_2}.
Moreover, the gradients of the bounded solutions \eqref{Regular_1:2_2} are unbounded around the poles.

Obviously, superpositions of solutions \eqref{Regular_1:2_1}, or  \eqref{Regular_1:2_2} with different
complex parameters $\omega$, running in some (discrete or continuous) sets in $\mathbb{C}_\omega$,
describe more general bounded solutions to the TAE with spin $1/2$.

One more remark. In the case $\sigma=+1$ the solutions \eqref{General_1:2} are unbounded
on the horizons $r_{\pm}$ due to the factor $\Delta(r)^{-1/2}$.
These {\em stationary} singularities cannot be removed by any choice of the function
${}_{\sigma\over 2}\mathfrak{A}\left(T_\sigma,W_\sigma\right)$, since it depends
on the two variables $T_\sigma$ and $W_\sigma$, not on the single one $r$.
The variables $t,r,\theta$ enter in $T_\sigma$ and the variables $\varphi,r,\theta$ enter in $W_\sigma$ in a complex way.
As a result, the variable $r$ cannot be disentangled from the the function
${}_{\sigma\over 2}\mathfrak{A}\left(T_\sigma,W_\sigma\right)$
and one is not able to compensate the singularity due to the factor $\Delta(r)^{-1/2}$ which does not depend
neither on the time $t$, nor on the angles $\varphi$ and $\theta$.

\section{Conclusion}
In the present paper, we have demonstrated that the confluent Heun functions are an adequate
and natural tool for a unified description of linear perturbations of gravitational field of the Kerr metric
outside the event horizon, as well as in interior domains. These functions
give us an effective tool for exact mathematical treatment of different boundary problems
and the corresponding physical phenomena.
They can help us to solve old mathematical issues, related to the Teukolsky separation
of the variables, as well as to study new physical problems.
The same approach works, too, for the Regge-Wheeler and  Zerilli equations in the Schwarzschild metric \cite{F, Fiziev2009b}.

Large classes of exact solutions to the perturbation equations of the Kerr metric were described
here for the first time. All possible types of solutions were classified uniformly
in terms of confluent Heun's functions and confluent Heun's polynomials, using their specific properties.
As we saw, the variety of the different solutions and possible spectra is much reacher than, for example,
the variety of the corresponding solutions and spectra of the Hydrogen problem in quantum mechanics,
solved exactly in terms of the confluent hypergeometric functions \cite{RS}.
Mathematically, this is obviously caused by the presence of one more regular singular point
in the confluent Heun equation.

We have to stress especially the newly obtained {\em singular} polynomial solutions to the Teukolsky
angular equation. They differ drastically from the well-known analogous {\em regular} polynomial solutions
to the quantum Hydrogen problem.
The singular polynomial solutions to the Teukolsky angular equation present an auxiliary mathematical construction
which seems suitable for simple and natural perturbative description of the collimation of radiated fields of all spins
$|s|>0$ in the Kerr metric, see \cite{PFDS, PFDS_astroph:HE}.

For spin $1/2$ we have proved that the singular kernels, constructed from polynomial solutions,
can produce bounded solutions of the continuous spectrum to the TME with very interesting physical properties:
These solutions describe collimated {\em one-way} running waves in the Teukolsky perturbation theory correctly.

For spin 1 we have also proved the existence of double polynomial solutions of the continuous spectrum to the TME
and can reach similar results in a more complicated way, since there we meet
a new physical phenomenon -- the electromagnetic superradiance \cite{Starobinskiy,Teukolsky,Chandra}.
For spin 2 the supperradiance is known to be quite stronger, but we have no continuous spectrum of the TME
and the problem needs special treatment.
We shall consider these two important cases separately.

One can hope that the collimated one-way running waves, cropping up for the first time in the present paper,
are able to describe the real astrophysical jets, observed at very different scales in the Universe.
This still speculative idea needs a more detailed mathematical development and
a careful confrontation with the real astrophysical observations.
It indicates the existence of a new universal mechanism for collimation of radiation of all spins $|s|>0$
by the pure gravitational field of rotating compact astrophysical objects of different nature.

\vskip .7truecm
\noindent{\bf \Large Acknowledgments}
\vskip .3truecm
I am thankful to Kostas Kokkotas, Luciano Rezzolla and Edward Malec
for the stimulating discussion  of exact solutions to the Regge-Wheeler and Teukolsky equations
and different boundary problems during the XXIV Spanish Relativity Meeting,
E.R.E. 2006, to Edward Malec for his kind invitation to visit
the Astrophysical Group of the Uniwersytet Jagiellonski, Crakow, Poland in May 2007
and to participants of the seminar there,
to participants of the Conference "Gravity, Astrophysics and Strings at the Black Sea" 2007,
Primorsko, Bulgaria,
to Goran Djeorjevic for his kind invitation to visit the Department of Physics, University of Nis,
Serbia in December 2007 and to give a talk there,
to Luciano Rezzolla -- for his kind invitation to visit the Albert Einstein Institute of Gravitational Physics in
Golm, Germany in March 2009 and to members of the Numerical Relativity Group there for numerous discussions.

The author is thankful to Denitsa Staicova and Roumen Borissov for numerous
discussions during the preparation of the present paper,
to Shahar Hod -- for his kind help in enriching the references, to
Jerome Gariel for drawing attention to the early papers by
Marcilhacy G and by Blandin J, Pons R, Marcilhacy G (see \cite{TME_Heun})
and for sending the corresponding copies, to Sam Dolan for drawing attention to the first paper
in \cite{HalfInteger} and for sending a copy of his PhD Thesis and to and to Sigbjorn Hervik
for drawing my attention to the article by Kayll Lake in [23]..

I would like to express my special gratitude to Professor Saul Teukolsky for his comments on the present paper
and his useful suggestions, as well as to Professor Alexey Starobinsky, to Professor George Alekseev,
to Professor Sergei Slavyanov, to Professor Irina Aref'eva and to Professor Igor Volovich for the useful
discussions and comments.

I am also thankful to the Bogolubov Laboratory of Theoretical Physics, JUNR, Dubna, Russia for the hospitality
and good working conditions during my stay there in the summer of 2009 and in February 2010.

This paper was supported by the Foundation "Theoretical and
Computational Physics and Astrophysics" and by the Bulgarian National Scientific Found
under contracts DO-1-872, DO-1-895 and DO-02-136.

\end{document}